\documentclass[twocolumn,twocolappendix]{aastex631}

\newcommand\com{\texttt{Commander} dust}
\newcommand\hi{H~\textsc{i}}
\newcommand\hipi{H\textsc{i}4PI}
\newcommand\hip{H~\textsc{i}-based polarization template}
\newcommand\kms{km~s$^{-1}$}

\newcommand{\nsideof}[1]{$N_{\rm side} = {#1}$}
\newcommand{\eqref}[1]{Equation~\ref{#1}}
\newcommand{\secref}[1]{Section~\ref{#1}}
\newcommand{\figref}[1]{Figure~\ref{#1}}

\newcommand{\eq}[1]{\begin{equation} #1 \end{equation}} 
\newcommand{\parens}[1]{\left ( #1 \right )} 
\newcommand{\nhat}{\mathbf{\hat{n}}}

\newcommand{\amin}{\(\stackrel{\:'}{\textstyle.}\)}
\newcommand{\adeg}{\(\stackrel{\:\circ}{\textstyle.\rule{0pt}{0.65ex}}\)}

\shorttitle{Filamentary Dust Polarization and the Morphology of \hi\ Structures}
\shortauthors{Halal et al.}

\begin{document}

\title{Filamentary Dust Polarization and the Morphology of Neutral Hydrogen Structures}

\correspondingauthor{George Halal}
\email{georgech@stanford.edu}

\author[0000-0003-2221-3018]{George Halal}
\affiliation{Department of Physics, Stanford University, Stanford, CA 94305, USA}
\affiliation{Kavli Institute for Particle Astrophysics and Cosmology (KIPAC), Stanford University, Stanford, CA 94305, USA}

\author[0000-0002-7633-3376]{Susan E. Clark}
\affiliation{Department of Physics, Stanford University, Stanford, CA 94305, USA}
\affiliation{Kavli Institute for Particle Astrophysics and Cosmology (KIPAC), Stanford University, Stanford, CA 94305, USA}

\author[0000-0002-7471-719X]{Ari Cukierman}
\affiliation{Department of Physics, Stanford University, Stanford, CA 94305, USA}
\affiliation{Kavli Institute for Particle Astrophysics and Cosmology (KIPAC), Stanford University, Stanford, CA 94305, USA}
\affiliation{SLAC National Accelerator Laboratory, Menlo Park, CA 94025, USA}
\affiliation{Department of Physics, California Institute of Technology, Pasadena, CA 91125, USA}

\author[0000-0003-0848-2756]{Dominic Beck}
\affiliation{Department of Physics, Stanford University, Stanford, CA 94305, USA}
\affiliation{Kavli Institute for Particle Astrophysics and Cosmology (KIPAC), Stanford University, Stanford, CA 94305, USA}
\affiliation{SLAC National Accelerator Laboratory, Menlo Park, CA 94025, USA}

\author{Chao-Lin Kuo}
\affiliation{Department of Physics, Stanford University, Stanford, CA 94305, USA}
\affiliation{Kavli Institute for Particle Astrophysics and Cosmology (KIPAC), Stanford University, Stanford, CA 94305, USA}
\affiliation{SLAC National Accelerator Laboratory, Menlo Park, CA 94025, USA}

\begin{abstract}
Filamentary structures in neutral hydrogen (\hi) emission are well aligned with the interstellar magnetic field, so~\hi\ emission morphology can be used to construct templates that strongly correlate with measurements of polarized thermal dust emission. We explore how the quantification of filament morphology affects this correlation. We introduce a new implementation of the Rolling Hough Transform (RHT) using spherical harmonic convolutions, which enables efficient quantification of filamentary structure on the sphere. We use this Spherical RHT algorithm along with a Hessian-based method to construct~\hip s. We discuss improvements to each algorithm relative to similar implementations in the literature and compare their outputs. By exploring the parameter space of filament morphologies with the Spherical RHT, we find that the most informative~\hi\ structures for modeling the magnetic field structure are the thinnest resolved filaments. For this reason, we find a~$\sim10\%$ enhancement in the~$B$-mode correlation with polarized dust emission with higher-resolution~\hi\ observations. We demonstrate that certain interstellar morphologies can produce parity-violating signatures, i.e., nonzero~$TB$ and~$EB$, even under the assumption that filaments are locally aligned with the magnetic field. Finally, we demonstrate that~$B$ modes from interstellar dust filaments are mostly affected by the topology of the filaments with respect to one another and their relative polarized intensities, whereas~$E$ modes are mostly sensitive to the shapes of individual filaments.
\end{abstract} 


\section{Motivation} \label{sec:intro}

Modeling polarized dust emission is crucial for studying various astrophysical phenomena in the interstellar medium (ISM) and for analyzing the polarization of the cosmic microwave background (CMB). Aspherical rotating dust grains preferentially align their short axes with the local magnetic field, resulting in their thermal emission being linearly polarized \citep{1975duun.book..155P}. Dust polarization thus traces the plane-of-sky magnetic field orientation and is widely used to trace magnetic field structure in the Galaxy \citep[e.g.,][]{Han:2017}. At large scales and frequencies greater than approximately 70 GHz, polarized dust emission is the predominant polarized CMB foreground \citep{2016A&A...594A..10P}. The accurate modeling and elimination of the dust contribution to CMB polarization measurements are essential to search for an excess polarization signal induced by primordial gravitational waves \citep{Kamionkowski_1997,Seljak_1997,Seljak__1997}.

Galactic neutral hydrogen (\hi) emission is a tracer of the neutral medium that can be fruitfully compared to the dust distribution. \hi\ and dust trace similar volumes of the diffuse ISM \citep{1996A&A...312..256B,2017ApJ...846...38L}. Much of the diffuse~\hi\ emission is organized into filamentary structures that show significant alignment with the plane-of-sky magnetic field orientation \citep{2014ApJ...789...82C,2015PhRvL.115x1302C}. The spectroscopic nature of~\hi\ measurements means that these structures can be studied in~3D, namely as a function of longitude, latitude, and radial velocity with respect to the local standard of rest $v_{\rm lsr}$, i.e., the Doppler-shifted frequency of the 21 cm line \citep{Clark:2018}. Furthermore, because~\hi\ and broadband thermal dust emission are independently observed, cross correlations between the two are free from correlated telescope systematics. \hi\ data are also not contaminated by the cosmic infrared background~\citep[][]{2019ApJ...870..120C}. 

Using these insights, \citet{ClarkHensley} developed a model of polarized dust emission based solely on~\hi\ measurements. Cross correlations between this dust polarization model and millimeter-wave polarization data have proven useful for characterizing dust properties such as the spectral index \citep{BKxHI}. \citet{ClarkHensley} used the Rolling Hough Transform~\citep[RHT;][]{2014ApJ...789...82C,2020ascl.soft03005C} algorithm as a first step for quantifying the orientations of linear dust filaments. The RHT has free parameters that set the scale and shape of the identified filaments. This is ideal for the exploration of different filament morphologies and their polarization effects. 

The RHT algorithm runs on images or flat-sky projections of small patches of the sky. It is possible to construct an~\hip\ on the full sky by projecting a small patch of the spherical map around each pixel to an image, running the algorithm, and projecting the result back to the sphere, as done in \citet{ClarkHensley} for a single set of parameters. However, this is computationally expensive to perform for multiple sets of parameters. In this paper, we develop an algorithm for running the RHT directly on the sphere using spherical harmonic convolutions.

Another filament-finding algorithm that can be used to construct polarization templates from~\hi\ emission is based on the Hessian matrix \citep[e.g.,][]{Cukierman}. We explore the advantages and disadvantages of the Hessian-based algorithm relative to the RHT-based algorithm. 

The paper is organized as follows. We introduce the dust emission and~\hi\ data used in this work in \secref{sec:data}. We explore how different modifications to the Hessian-based polarization template affect the correlation with polarized dust emission in \secref{sec:imp}. We introduce a spherical convolution version of the RHT in \secref{sec:sphericalrht}. We use it to explore the polarization effects of different filament morphologies and how filament-finding algorithms can be used for determining morphologies that produce parity-violating polarization signatures in \secref{sec:filmorph}. We summarize and conclude in \secref{sec:sum}.

\section{Data} \label{sec:data}
\subsection{Dust Emission\label{subsec:dustdata}}
We make use of two sets of Stokes~$I$,~$Q$, and~$U$ Planck data products at~353~GHz provided by the Planck Legacy Archive\footnote{\url{pla.esac.esa.int}}. The first is the set of Planck \com\ maps with an angular resolution of~$5'$, constructed by component separation applied to the Planck frequency maps \citep{refId0}. The second is the set of~353~GHz maps from Planck data release R3.01 with an angular resolution of~5$'$ \citep{2020A&A...641A...1P}. While the former is processed to remove emission other than dust, the latter contains contributions from multiple components. We compare these two data products in Section~\ref{subsec:dustmask} and use the \com\ maps for all subsequent analyses. The reported results are insensitive to smoothing the Planck data, so we use them at their native resolution.

For cross-spectrum calculations between these maps and other polarization data products in this paper, we use the full-mission maps. When calculating autospectra of these maps, we compute cross spectra of the half-mission splits to avoid noise bias.

For most of the analysis in this paper, we use the Planck~70\% sky fraction Galactic plane mask \citep{2015A&A...576A.104P}. However, in \secref{subsec:dustmask}, we also employ the 20\%, 40\%, 60\%, and 80\% sky fraction Galactic plane masks. The higher the sky fraction, the greater the contribution from lower Galactic latitudes.

\subsection{Neutral Hydrogen\label{subsec:hidata}}
For the Galactic neutral hydrogen (\hi) emission, we use the~\hi~4$\pi$ Survey \citep[\hipi;][]{2016A&A...594A.116H}, which has the highest-resolution full-sky measurements of the~21~cm hyperfine transition to date. \hipi\ merges data from the Effelsberg-Bonn~\hi\ Survey \citep[EBHIS;][]{2016A&A...585A..41W} and the Parkes Galactic All-Sky Survey \citep[GASS;][]{2009ApJS..181..398M} to achieve an angular resolution of~16\amin2, a spectral resolution of~1.49~\kms, and a normalized brightness temperature noise of~$\sim53$~mK for a 1~\kms\ velocity channel. We use the publicly available~\hi\ intensity data described in \citet{ClarkHensley}, binned into velocity channels of equal integrated intensity in each pair of channels moving symmetrically outward from the local standard of rest.

In \secref{subsec:galfa}, we also use~\hi\ emission data from the Galactic Arecibo L-Band Feed Array~\hi\ Survey \citep[GALFA-\hi;][]{2018ApJS..234....2P}, which is higher resolution but only covers~$\sim32\%$ of the sky. GALFA-\hi\ has an angular resolution of~4\amin1, a spectral resolution of~0.184~\kms, and a normalized brightness temperature noise of~150~mK for a 1~\kms\ channel. We also use the publicly available~\hi\ intensity data described in \citet{ClarkHensley}, binned into velocity channels of equal width of 3.7~\kms. These maps span the range~1\adeg5~$<~\mathrm{decl.}~<$35\adeg5 to avoid telescope scan artifacts at the edges of the Arecibo declination range.

\section{\hi-based Dust Polarization Prediction} \label{sec:hip}
The~\hip s are constructed by measuring the orientation of linear structures to determine the polarization angle and combining this information with some weighting representing the polarized intensity at different locations in the map. The orientation can be determined by different algorithms applied to the~\hi\ intensity maps. In this section, we summarize the two algorithms we use in this paper for polarization angle determination and describe how their outputs are used along with different polarized intensity weighting schemes to construct~\hip s.

\subsection{RHT-based Angle Determination\label{subsec:rht}}
The RHT is a computer vision algorithm that identifies linear structures and their orientations in images \citep{2014ApJ...789...82C,2020ascl.soft03005C}. The steps involved in this process are:
\begin{enumerate}
    \item Unsharp masking, which involves subtracting a version of the map smoothed to a given scale,~$\theta_{\rm FWHM}$, from the original map. This step acts as a high-pass filter of the map to remove larger-scale emission.
    \item Bit masking, which converts all pixels with negative values to zero and all pixels with positive values to one.
    \item Applying the Hough transform \citep{osti_4746348} on a circular window of a given diameter,~$D_{\rm W}$, centered on each pixel to quantify the relative intensities of differently oriented linear structures passing through that pixel.
    \item Storing only the linear intensities over a certain threshold fraction,~$Z$, of the window diameter. The output is stored as linear intensity as a function of orientation~$R(\nhat, \theta, v)$.
\end{enumerate}
This algorithm, therefore, has three free parameters, $\theta_{\rm FWHM}$, $D_{\rm W}$, and~$Z$, which can be tuned to different values for different applications.

\subsection{Hessian-based Angle Determination} \label{subsec:hess}
Hessian-based filament identification has been applied to different maps, e.g., Planck 353 GHz total intensity maps \citep{2016A&A...586A.135P,Planck2016XXXVIII}, \hipi\ \hi\ intensity and Planck 857 GHz total intensity maps \citep{Kalberla2021}, Herschel images of molecular clouds \citep{2013ApJ...777L..33P}, and simulations of the cosmic web \citep{2000PhRvL..85.5515C,2009MNRAS.396.1815F}. In this work, we use the version of the Hessian-based filament-finding algorithm described in \citet{Cukierman}.

The Hessian matrix serves as a tool to determine the orientation of filaments. It contains information about the local second derivatives. A negative curvature indicates the presence of at least one negative Hessian eigenvalue. By examining a map for areas exhibiting negative curvature, we can identify possible filaments.

We apply the Hessian to the~\hi\ intensity maps in individual velocity bins~$I$. We work in spherical coordinates with polar angle~$\theta$ and azimuthal angle~$\phi$. The local Hessian matrix is given by 
\eq{ H \equiv \parens{ \begin{array}{cc} H_{xx} & H_{xy} \\
							    H_{yx} & H_{yy} \end{array} } , }
where		    
\begin{eqnarray}
		 H_{xx}  & = & \frac{\partial^2 I}{\partial \theta^2} ,  \\
		 H_{yy} & = & \frac{1}{\sin^2 \theta} \frac{\partial^2 I}{\partial \phi^2},  \\ 
		 H_{xy} = H_{yx} & = & - \frac{1}{\sin\theta} \frac{\partial^2 I}{\partial \phi \partial \theta} . 
\end{eqnarray}
The eigenvalues are
\eq{ \lambda_{\pm} = \frac{1}{2} \parens{ H_{xx} + H_{yy} \pm \alpha } , }
where
\eq{ \alpha \equiv \sqrt{ \parens{ H_{xx} - H_{yy} }^2 + 4 H_{xy}^2 } . }

For the local curvature to be negative along at least one axis, we require~$\lambda_{-} < 0$. We also require~$\lambda_{-}$to be the larger of the two eigenvalues in magnitude such that this negative curvature is the dominant local morphology. In constructing \hip s, we define a weighting~$w_{\rm H}$ for each pixel at each velocity that is equal to~$\lambda_{-}$ when the eigenvalues satisfy our two requirements and equal to zero when they do not.

The orientation of the filaments is determined by the local eigenbasis. The polarization angle is determined as
\eq{ \theta_{\rm H} = \arctan \parens{ \frac{H_{xx} - H_{yy} + \alpha}{2 H_{xy}} } . \label{eq:thetav} }

\subsection{\hi-based Polarization Template Construction} \label{subsec:hipcons}
The polarization angle determined using the RHT and Hessian algorithms can be combined with some weighting representing the local contribution to the \hi-based polarized intensity to construct~\hip s. For instance, \citet{ClarkHensley} normalize the RHT-measured linear intensity $R(\nhat, \theta, v)$ over different orientation bins such that
\eq{\sum_\theta R(\nhat, \theta, v) = 1. \label{eq:norm}}
They use the normalized~$R(\nhat, \theta, v)$ and the~\hi\ intensity maps~$I_{\rm HI}(\nhat, v)$ as the weighting to produce Stokes~$Q_{\rm RHT}$ and~$U_{\rm RHT}$ maps as 
\eq{Q_{\rm RHT}(\nhat, v) = I_{\rm HI}(\nhat, v) \sum_\theta R(\nhat, \theta, v) \cos{2\theta}, \label{eq:QRHT}}
\eq{U_{\rm RHT}(\nhat, v) = I_{\rm HI}(\nhat, v) \sum_\theta R(\nhat, \theta, v) \sin{2\theta}. \label{eq:URHT}}
These maps have the same units (K~\kms) as the intensity maps, $I_{\rm HI}(\nhat, v)$. To construct the~\hip, the Stokes parameter maps are integrated over velocity channels as
\eq{Q_{\rm RHT}(\nhat) = \sum_v Q_{\rm RHT}(\nhat, v), \label{eq:QRHTsum}}
\eq{U_{\rm RHT}(\nhat) = \sum_v U_{\rm RHT}(\nhat, v). \label{eq:URHTsum}}

\begin{figure}[t!]
\includegraphics[width=\columnwidth]{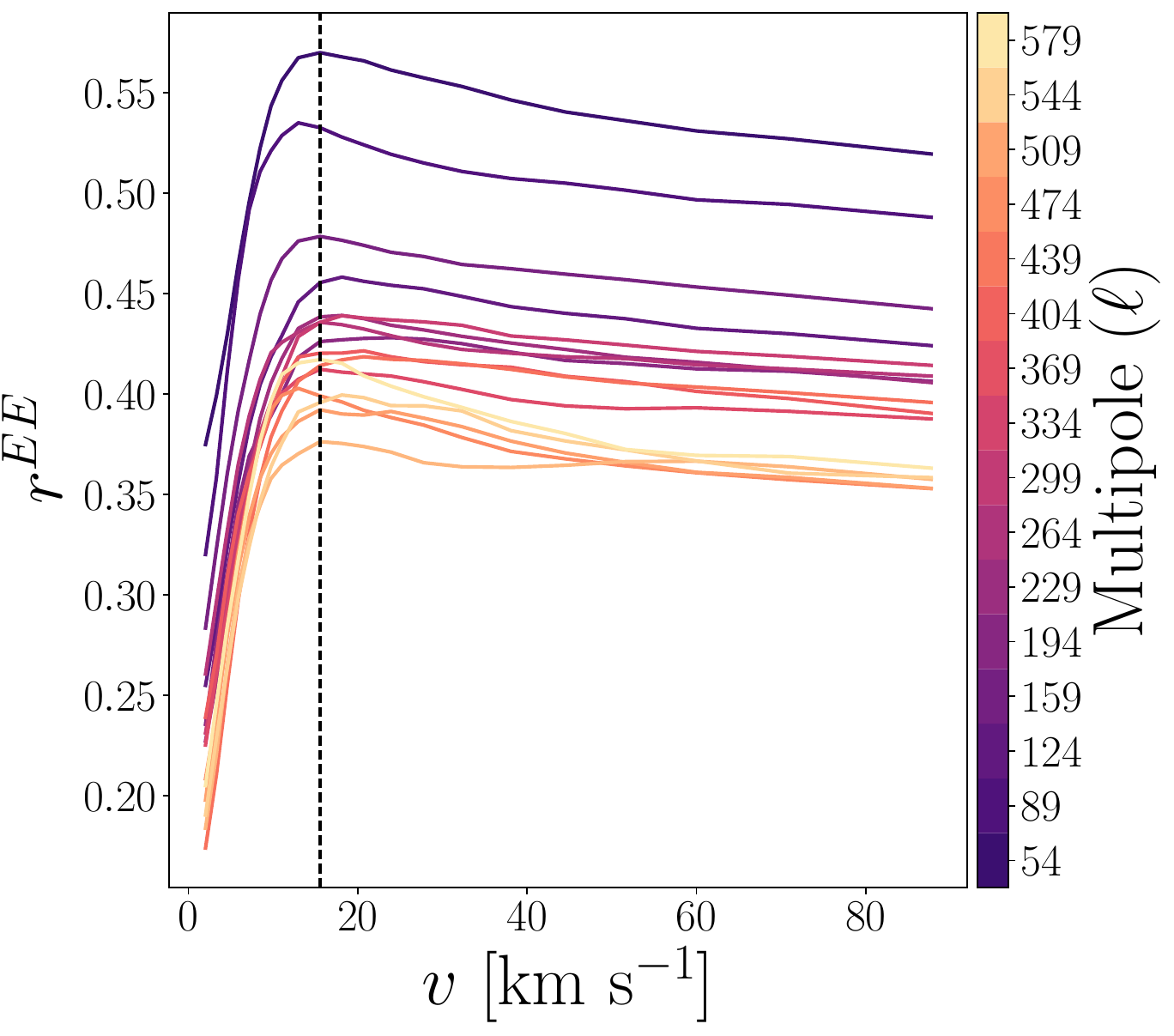}
\caption{The $EE$ correlation ratio on the Planck~70\% sky fraction Galactic plane mask of the Planck \com\ maps with~\hip s produced by the Hessian algorithm applied to~\hipi\ data. The leftmost point is for the~\hip\ at~2.03~\kms, and each successive point on each curve corresponds to the addition of information from the two adjacent velocity channels in the positive and negative directions. The labels on the horizontal axis correspond to the velocity centers of the positive velocity channels being added. The different curves correspond to different multipole bins shown in the color bar. The vertical dashed line corresponds to the integrated~\hip\ over the velocity range~$-13$~\kms~$<~v_{\rm lsr}~<~16$~\kms, after which the correlation saturates and starts decreasing over most of the multipole bins considered as information from more velocity channels is added.
\label{fig:vel}}
\end{figure}

The Hessian-based filament-finding algorithm (hereafter Hessian algorithm), by contrast, uses the local eigenbasis of the Hessian matrix to determine the orientation of linear structures, and the negative eigenvalues to determine the Stokes weighting \citep{Cukierman}. See \secref{subsec:hess} for details. Equations~\ref{eq:QRHT} and \ref{eq:URHT}, therefore, become 
\eq{Q_{\rm H}(\nhat, v) = w_{\rm H}(\nhat, v) \cos{2\theta_{\rm H}(\nhat, v)}, \label{eq:QHess}}
\eq{U_{\rm H}(\nhat, v) = w_{\rm H}(\nhat, v) \sin{2\theta_{\rm H}(\nhat, v)}, \label{eq:UHess}}
where $\theta_{\rm H}$ is the polarization angle perpendicular to the orientation of the local linear structure determined by the Hessian, and $w_{\rm H}$ is formed from the negative eigenvalues as described in \secref{subsec:hess}. These maps are then summed over velocity as in Equations \ref{eq:QRHTsum} and \ref{eq:URHTsum} to produce $Q_{\rm H}(\nhat)$ and $U_{\rm H}(\nhat)$.

Although work conducted across extensive portions of the high-Galactic latitude sky indicates that there could be a minor uniform misalignment between the filaments and the orientation of the magnetic field as measured by Planck, the angle of misalignment amounts to approximately~$\sim2^\circ$-$5^\circ$ only \citep[][]{Huffenberger_2020,Clark2021,Cukierman}. Rotating the~\hip\ angles by this amount to emulate this misalignment effect leads to only a slight enhancement in the correlation at the level of~$\sim$0.1\%-0.5\% \citep{Cukierman}, and we do not apply this rotation here.

\section{Improvements in~\hi-based Dust Polarization Prediction} \label{sec:imp}

In this section, we employ the Hessian algorithm on~\hipi\ and GALFA-\hi~data to construct~\hip s. Working within the Hessian-based framework, we examine how to construct templates that correlate most strongly with Planck polarized dust emission maps. We later contrast these templates with alternative maps based on the Spherical RHT.

\subsection{Velocity Selection \label{subsec:velsel}}

Since we expect the correlation between~\hi\ and dust to vanish for high-velocity clouds \citep{1986A&A...170...84W,2011A&A...536A..24P,2017ApJ...846...38L}, we restrict the velocity range over which we integrate the~\hip\ in \figref{fig:vel}. More generally, any contribution to the~\hip\ from noise, data artifacts, or~\hi\ emission that is not correlated with dust structure will tend to decrease the measured correlation between the template and the polarized dust emission. To restrict the velocity range, we calculate the correlation ratio defined as
\eq{
r_\ell^{{\rm data} \times {\rm HI}} = \frac{D_\ell^{{\rm fm} \times {\rm HI}}}{
\sqrt{D_\ell^{{\rm hm1} \times {\rm hm2}} \times D_\ell^{{\rm HI} \times {\rm HI}}}}, \label{eq:corr_rat}
}
where~$D_\ell^{m_1 \times m_2}$ is the cross-spectrum bandpower between two maps,~$m_1$ and~$m_2$, in the multipole bin~$\ell$. All power spectra in this paper are computed with the \texttt{pspy}\footnote{\url{https://github.com/simonsobs/pspy}} code \citep{Louis:2020}. We use the Planck \com\ maps as the \textit{data}, where the full-mission maps (\textit{fm}) are used for the cross spectra with the~\hip\ and the half-mission splits (\textit{hm1} and \textit{hm2}) are used in the denominator. The~\textit{HI} in this equation refers to the~\hip, which in this case is constructed using the Hessian algorithm on the~\hipi\ dataset.

\begin{figure}[t!]
\includegraphics[width=\columnwidth]{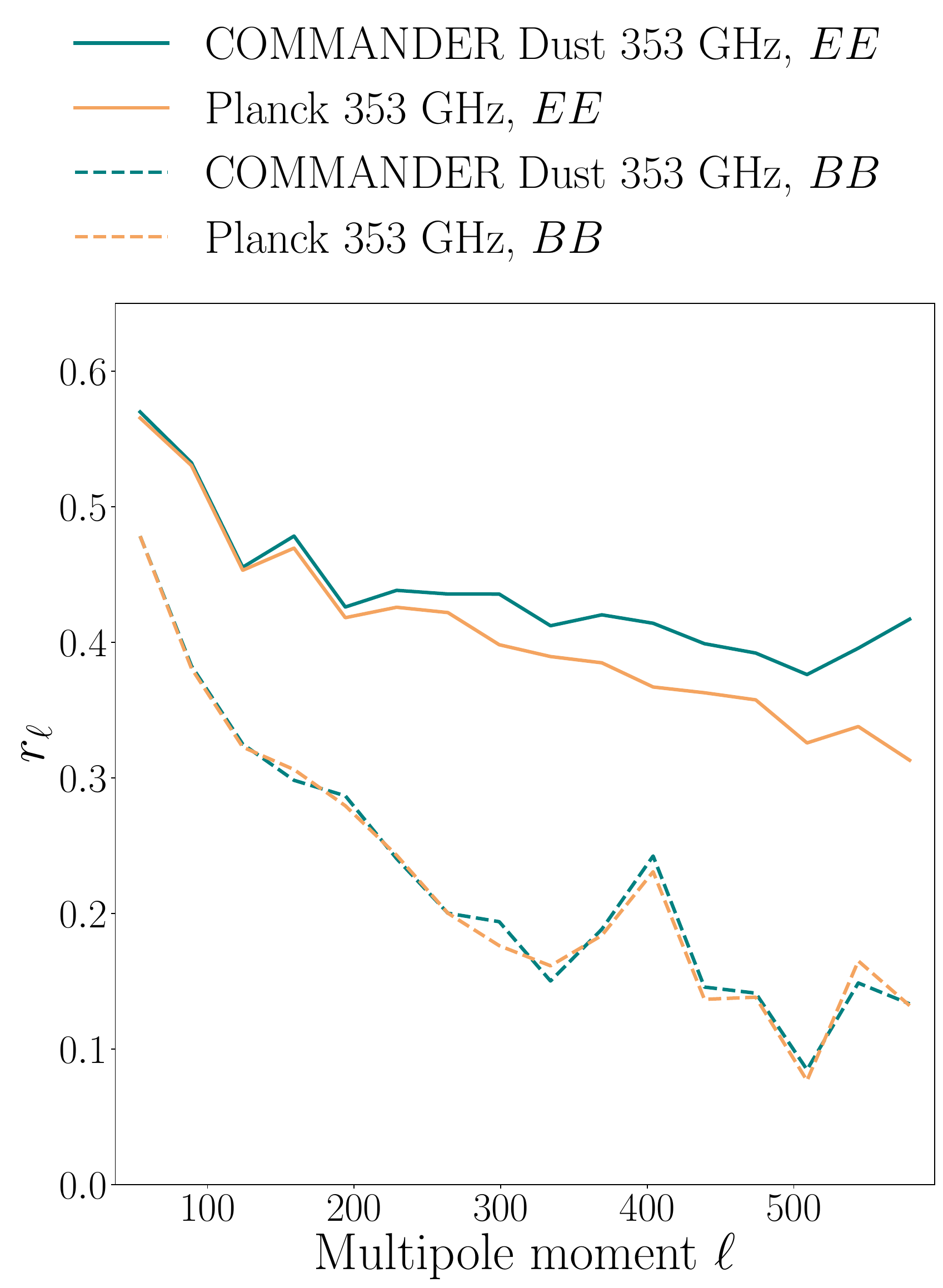}
\caption{Comparison of the~$EE$ (solid) and~$BB$ (dashed) correlation ratios on the Planck~70\% sky fraction Galactic plane mask of the Planck \com\ maps (teal) and the Planck frequency maps (sandy brown) at 353~GHz with the~\hip\ constructed from applying the Hessian algorithm to~\hipi\ data at each velocity channel and integrating the resulting maps over the velocity range~$-13$~\kms~$<~v_{\rm lsr}~<~16$~\kms.
\label{fig:cvp}}
\end{figure}

\begin{figure*}[t!]
\includegraphics[width=2\columnwidth]{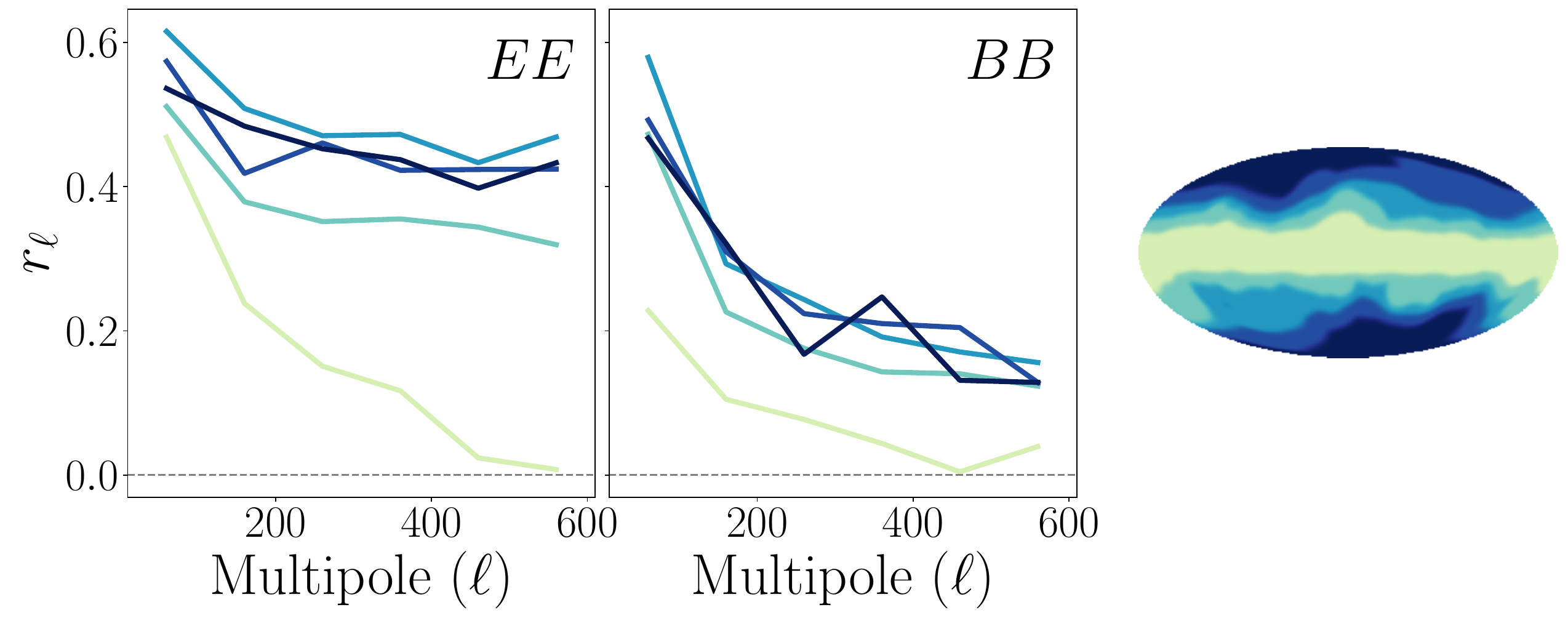}
\caption{The~$EE$ (left) and~$BB$ (middle) correlation ratios of the Planck \com\ maps with the~\hip\ constructed using the Hessian algorithm on the~\hipi\ intensity maps over the different non-overlapping masks shown on the right. The masks are the Planck~20\% sky fraction Galactic plane mask (darkest), the inverted Planck~80\% sky fraction Galactic plane mask (lightest), and the differences between the Planck~20\%, 40\%, 60\%, and 80\% sky fraction Galactic plane masks (other shades of blue) shown in a Mollweide projection in Galactic coordinates centered on the Galactic center.
\label{fig:permask}}
\end{figure*}

The linear polarization field described by the Stokes~$Q$ and~$U$ maps can be decomposed into~$E$-mode and~$B$-mode components \citep{Seljak_1997,Zaldarriaga:2001}. In \eqref{eq:corr_rat},~\textit{data}$\times$\textit{HI} can be the correlation of any combination of the~$E$-mode, $B$-mode, or intensity components of the Planck \com\ maps and the~\hip s. For example, in \figref{fig:vel}, we calculate \eqref{eq:corr_rat} for their~$E$-mode components. We do not show the~$BB$ correlation ratio because it exhibits similar behavior. We use the Planck 70\% sky fraction Galactic plane mask for the correlation ratio calculations in this figure. We use the same multipole binning used in~\citet{BKxHI} for a direct comparison of results and because we did not find that multipoles higher than~600 provide additional insights. We start with the~\hip\ of the individual velocity channel at~$v_{\rm lsr}~=~2.03$~\kms. We pick this velocity channel because it has the highest~\hi\ intensity integrated over the unmasked sky. We plot the correlation ratio as a function of velocity integration range and spatial scale in \figref{fig:vel}. The x-axis on this figure is cumulative, i.e., moving toward higher velocities on these plots corresponds to symmetrically integrating outwards in the positive and negative directions from the starting velocity channel, adding one~\hip\ from each direction to the previous~\hip. 

The correlation ratio in all multipole bins saturates and even starts decreasing as information from more channel maps is added after a certain velocity. We conclude from this analysis that the~\hip\ is most strongly correlated with the polarized dust emission in the range~$-13$~\kms~$<~v_{\rm lsr}~<~16$~\kms\ over most of the multipole bins considered. We find the same range for~$B$ modes as well. We use this cut for the rest of the analysis in this paper. Note that the correlation is already at the $\sim$20\%-40\% level for the~\hip\ of the individual velocity channel at~$v_{\rm lsr}~=~2.03$~\kms. 

\citet{Cukierman} restrict the velocity range to~$-15$~\kms~$<~v_{\rm lsr}~<~4$~\kms. With our velocity selection, we achieve an additional~$\sim5\%$ increase in the~$EE$ and~$BB$ correlation ratios with the Planck \com\ maps relative to \citet{Cukierman}. \citet{2020ApJ...902..120P} propose the range~$-12$~\kms~$<~v_{\rm lsr}~<~10$~\kms\ for low-velocity clouds~(LVCs). These are the 1st and~99th percentiles of the velocity distribution of clouds with~\hi\ column density $N_{\rm HI} > 2.5 \times 10^{20}~{\rm cm}^{-2}$ located in the Northern and Southern Galactic Polar regions. The velocity selection we make is close to this range. A benefit of restricting our analysis to this velocity range is that our template is less likely to include contributions from gas at very different distances, which decreases the likelihood of mixing different physical scales.

Using the RHT algorithm for determining the polarization angle, the~\hi\ intensity maps as the weighting (see \secref{sec:imp}), and the Spearman rank correlation coefficient and mean angle alignment as the correlation metrics, \citet{ClarkHensley} did not see the decrease in the correlation after a certain velocity that we see in \figref{fig:vel}. The correlation asymptotes instead. There is also evidence that the intermediate velocity cloud (IVC) gas is organized into filaments that are aligned with their local magnetic fields \citep{Panopoulou:2019, 2021A&A...647A..16P}. The difference could be caused by the Hessian algorithm being more sensitive than the RHT to artifacts in low-signal velocity channels. We explore this in \secref{subsec:galfa}.

\begin{figure*}[t!]
\includegraphics[width=2\columnwidth]{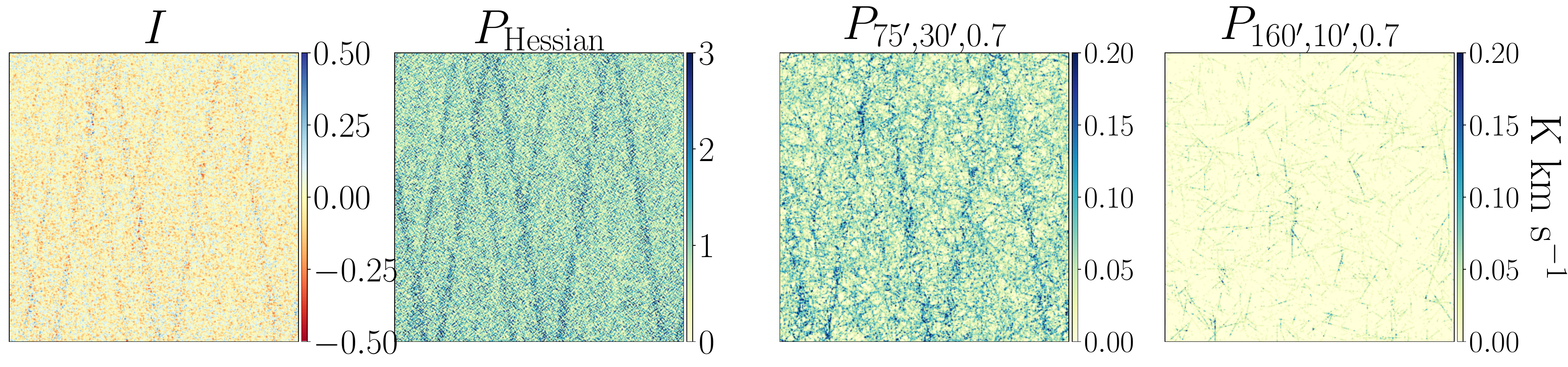}
\caption{Maps of a~$15^\circ \times 15^\circ$ patch of sky centered on R.A.~=~7\adeg5, decl.~=~28\adeg1 and~$v_{\rm lsr}~=~400.1$~\kms\ of the input intensity channel from GALFA-\hi\ with width~0.74~\kms\ (left) and polarized intensity of~\hip s constructed using the Hessian (middle left) and RHT (middle right and right) algorithms applied to this channel. The three parameters listed in the titles of the polarized intensity maps produced with the RHT algorithm are~$D_{\rm W}$,~$\theta_{\rm FWHM}$, and~$Z$, respectively, explained in~\secref{subsec:rht}.
\label{fig:galfa400}}
\end{figure*}

\subsection{Dust Map and Mask Comparisons \label{subsec:dustmask}}
We examine how the choice of dust emission maps and sky masks affects the correlation with the~\hip\ constructed using the Hessian algorithm on the aforementioned velocity selection in~\hipi\ data. We compare the~$E$- and~$B$-mode correlation ratios with the~\hip\ on the~70\% sky fraction mask between the Planck \com\ maps at~353~GHz and the Planck frequency maps at 353~GHz in \figref{fig:cvp}. We note that the difference is negligible in~$B$ modes and at low multipoles in~$E$ modes. The \com\ maps correlate more strongly than the Planck frequency maps at higher multipoles in~$E$ modes. This is due to the~CMB~$E$ modes, which contribute~$\sim10$\% of the~$E$-mode power to the 353 GHz frequency maps at~$\ell~>~300$. We therefore use the \com\ maps for the rest of the analysis in this paper.

We test how the correlation ratio between the Planck \com\ maps and the~\hip\ changes at different Galactic latitudes by utilizing the Planck sky fraction masks mentioned in \secref{sec:data}. We invert the~80\% sky fraction Galactic plane mask by switching ones to zeros, and vice versa, and call this the low-Galactic-latitude~20\% mask. We analyze the results for the high-Galactic latitude~20\% sky fraction mask, the low-Galactic-latitude~20\% mask, and the differences between the~20\% and~40\%,~40\% and~60\%, and~60\% and~80\% sky fraction Galactic plane masks. These five masks are shown in \figref{fig:permask}, along with the~$EE$ and~$BB$ correlation ratios calculated over these masks. Although a significant portion of the dust column stops being traced by~\hi\ at lower Galactic latitudes because it is associated with molecular gas there \citep{2017ApJ...846...38L}, we find a~$\mathbf{\sim20\%}$ correlation with the low-Galactic-latitude~20\% mask up to multipoles of~$\ell\sim400$. Since the velocity selection was optimized for the~70\% sky fraction Galactic plane mask, we test whether the reported correlations calculated with the low-Galactic-latitude~20\% mask increase when the~\hip s are integrated over a wider velocity range. We find that the correlation steadily increases over the entire multipole range considered with each template added out to~$\pm~90$~\kms, reaching~$\sim68\%$ at~$\ell\sim50$ in~$EE$. This increase is expected because our initial velocity selection includes less than~47\% of the total Galactic \hi\ column density in this mask.

\subsection{Effects of Resolution and Data Artifacts \label{subsec:galfa}}

In the previous subsections, we have only applied the Hessian to~\hipi\ data, which has an angular resolution of~16\amin2, using~\nsideof{1024}. In this subsection, we apply the Hessian algorithm to GALFA-\hi\ data, which has a much finer angular resolution of~4$'$, using~\nsideof{2048}. The velocity binning between the~\hipi\ and GALFA-\hi\ datasets is different, so we use the closest velocity range we can define for the GALFA-\hi\ dataset to the velocity selection we found using the~\hipi\ dataset, which is $-15$~\kms~$<~v_{\rm lsr}~<~18$~\kms. However, as shown in \figref{fig:vel}, most of the correlation comes from the~\hi\ emission near the local standard of rest, and we find the difference in dust correlation between the velocity ranges $-15$~\kms~$<~v_{\rm lsr}~<~18$~\kms\ and $-11$~\kms~$<~v_{\rm lsr}~<~15$~\kms\ to be negligible. 

\begin{figure*}[t!]
\includegraphics[width=2\columnwidth]{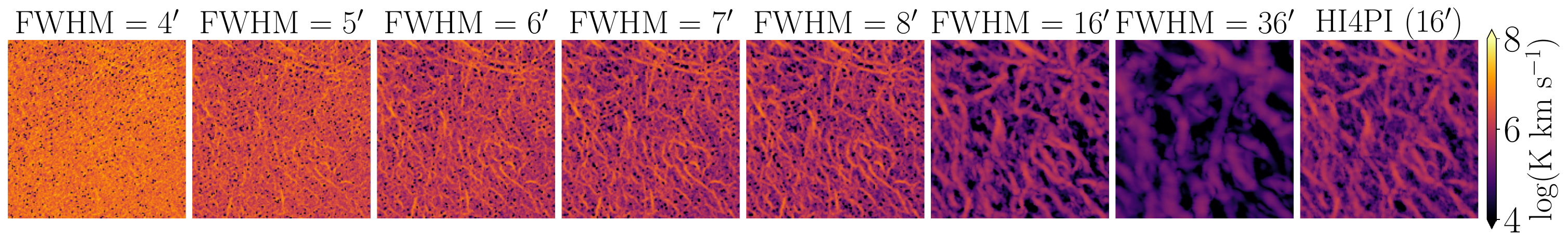}
\caption{Polarized intensity map projections of a~$400' \times 400'$ patch of sky, centered at $(l,~b)~=~(15^\circ,~50^\circ)$, of~\hip s constructed using the Hessian algorithm applied to GALFA-\hi\ intensity maps smoothed to different resolutions. The intensity map corresponding to the leftmost polarized intensity map is not smoothed, i.e., it has the native resolution of GALFA-\hi. The intensity maps corresponding to the polarized intensity maps to the right of the first map are smoothed with Gaussian kernels to the resolution stated in their titles. The last map on the right corresponds to the Hessian algorithm applied to~\hipi\ data at its native resolution, though it is integrated over a slightly different velocity range as mentioned in \secref{subsec:galfa}.
\label{fig:sm_maps}}
\end{figure*}

We find that the Hessian algorithm highlights any structure with significant local curvature, which includes scan-pattern artifacts and other emission that is irrelevant to the physical gas filament distribution. To demonstrate this, we apply the Hessian method to a high-velocity channel from GALFA-\hi\ centered on~$v_{\rm lsr}~=~400.1$~\kms, which has relatively little emission above the noise level in the region analyzed. We compare the polarized intensity maps of an example patch of sky of~\hip s constructed using the Hessian and RHT algorithms applied to this channel in \figref{fig:galfa400}. Polarized intensity is defined as
\eq{P = \sqrt{Q^2 + U^2}, \label{eq:polint}}
where~$Q$ and~$U$ are the Stokes parameter maps of the~\hip\ at~$v_{\rm lsr}~=~400.1$~\kms\ in this case. The RHT algorithm is discussed in \ref{subsec:rht}. The input intensity map in this figure clearly shows the scan-pattern artifacts. The same artifacts are also obvious in the polarized intensity maps of the Hessian-based template and the first RHT-based template. However, they become much less obvious in the second RHT-based template with different parameters. This shows a limitation of the Hessian method, which the RHT algorithm can be tuned to avoid. This supports the hypothesis made in \secref{subsec:velsel} concerning the decrease in the correlation with polarized dust emission when incorporating higher velocity channels beyond a specific threshold, which is observed only when using the Hessian method (\figref{fig:vel}) but not when using the RHT algorithm \citep{ClarkHensley}.

To mitigate this limitation, we experimented with various sensitivity-based weighting schemes for downweighting the low-intensity map pixels in each~\hi\ channel. We apply the Hessian algorithm to these new maps and find this weighting to have a negligible effect on the correlation ratio with polarized dust emission. This is true when using both GALFA-\hi\ and~\hipi\ data for the~\hi\ intensity. This implies that the low-intensity~\hi\ pixels do not strongly affect the template, i.e., it is not necessarily the low-intensity pixels that contain emission that is uninformative about filament orientations. Rather, this uninformative emission is likely scan-pattern artifacts and other data systematics.

\begin{figure*}[t!]
\includegraphics[width=2\columnwidth]{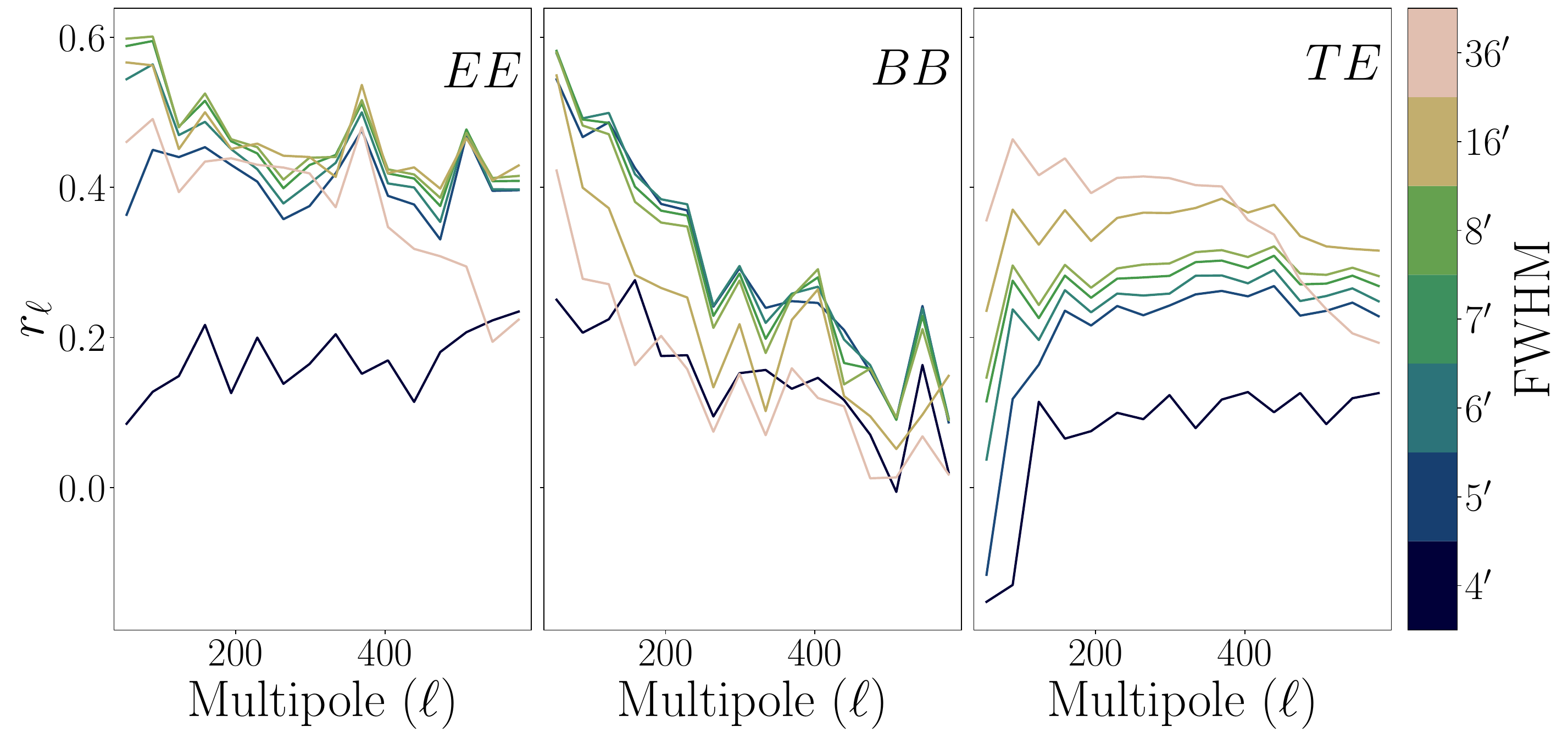}
\caption{The~$EE$ (left), $BB$ (middle), and~$TE$ (right) correlation ratios between the maps in \figref{fig:sm_maps} and the Planck \com\ maps at~353~GHz. The~$TE$ correlation ratio measures the correlation between the Planck total intensity and the templates'~$E$ modes. Correlations are computed on a combination of the GALFA-\hi\ and Planck 70\% sky fraction masks.
\label{fig:sm_specs}}
\end{figure*}

To modify the scale of structure that the Hessian algorithm is most sensitive to, we apply a Gaussian smoothing kernel to each of the GALFA-\hi\ intensity maps before applying the Hessian algorithm. \figref{fig:sm_maps} shows the effect of smoothing before applying the Hessian. The maps shown are projections of the polarized intensity over the velocity selection described in \secref{subsec:velsel} of an example patch of sky. The first map on the left corresponds to the Hessian algorithm run on the GALFA-\hi\ intensity maps at their native resolution of~$4'$. 
Note that it is difficult to see the filamentary structure because the Hessian algorithm is sensitive to other local variations, such as noise and scan-pattern artifacts. Each of the maps to the right of the first one corresponds to the Hessian algorithm applied to the GALFA-\hi\ intensity maps smoothed with a Gaussian kernel to the labeled FWHM resolution. Increasing the FWHM emphasizes real filamentary structure but also makes the filaments wider. We include maps for FWHM=$16'$ and~$36'$, which are the native resolutions of~\hipi\ and the Leiden/Argentine/Bonn (LAB) surveys \citep{Kalberla_2005}, respectively. We include the projection of the polarized intensity of an example patch of the sky when the Hessian algorithm is applied to~\hipi\ data at their native resolution of~16$'$ on the right of the figure for comparison with the map titled FWHM=$16'$. The velocity binning and the velocity range over which the templates of the two datasets are integrated are not identical, so we do not expect their resulting polarized intensity maps to be identical. When smoothed to the~\hipi\ beam, the GALFA-\hi\ maps are modestly more sensitive.

Smoothing the~\hi\ data before constructing the~\hip s deemphasizes small-scale noise at the cost of sensitivity to real small-scale~\hi\ structure that may correlate well with the measured polarized dust emission. We explore this trade-off in GALFA-\hi\ data by calculating the~$EE$,~$BB$, and~$TE$ correlation ratios of the different maps in \figref{fig:sm_maps} with the Planck \com\ maps and plot the results in \figref{fig:sm_specs}. For the~$TE$ case, we correlate the Planck dust total intensity with the templates'~$E$ modes. 

The trend is not consistent between the three panels. The~$TE$ correlation simply increases as the~GALFA-\hi\ data are smoothed, with an expected dip at the smoothing scale, which is only within the multipole range considered for the FWHM=36$'$ case. However, the~$EE$ and~$BB$ correlations are maximized when the~\hi\ data are smoothed to intermediate resolutions. While an increase in the~$EE$ and~$BB$ correlation ratios is what we should aim for, an increase in the~$TE$ correlation ratio is not necessarily better since we expect the real~$TE$ correlation ratio to be~$\sim0.36$ \citep{PlanckCollaboration:2020}. Therefore, a near-ideal~$E$-mode template should correlate with the Planck dust total intensity at about that level. However, in all cases, we achieve a significant improvement in the correlation with polarized dust emission by smoothing the map before applying the Hessian algorithm. The~\hip\ based on~\hipi\ data at their native~16$'$ resolution is similarly correlated with the dust polarization to the template based on the GALFA-\hi\ data smoothed to 16$'$. The map smoothed to 36$'$ decreases both the~$EE$ and~$BB$ correlations with the Planck \com\ maps. This shows the utility of the higher-resolution~\hi\ intensity data from the GALFA-\hi\ and~\hipi\ surveys in modeling polarized dust emission over the lower-resolution LAB survey.

\begin{figure}[t!]
\includegraphics[width=\columnwidth]{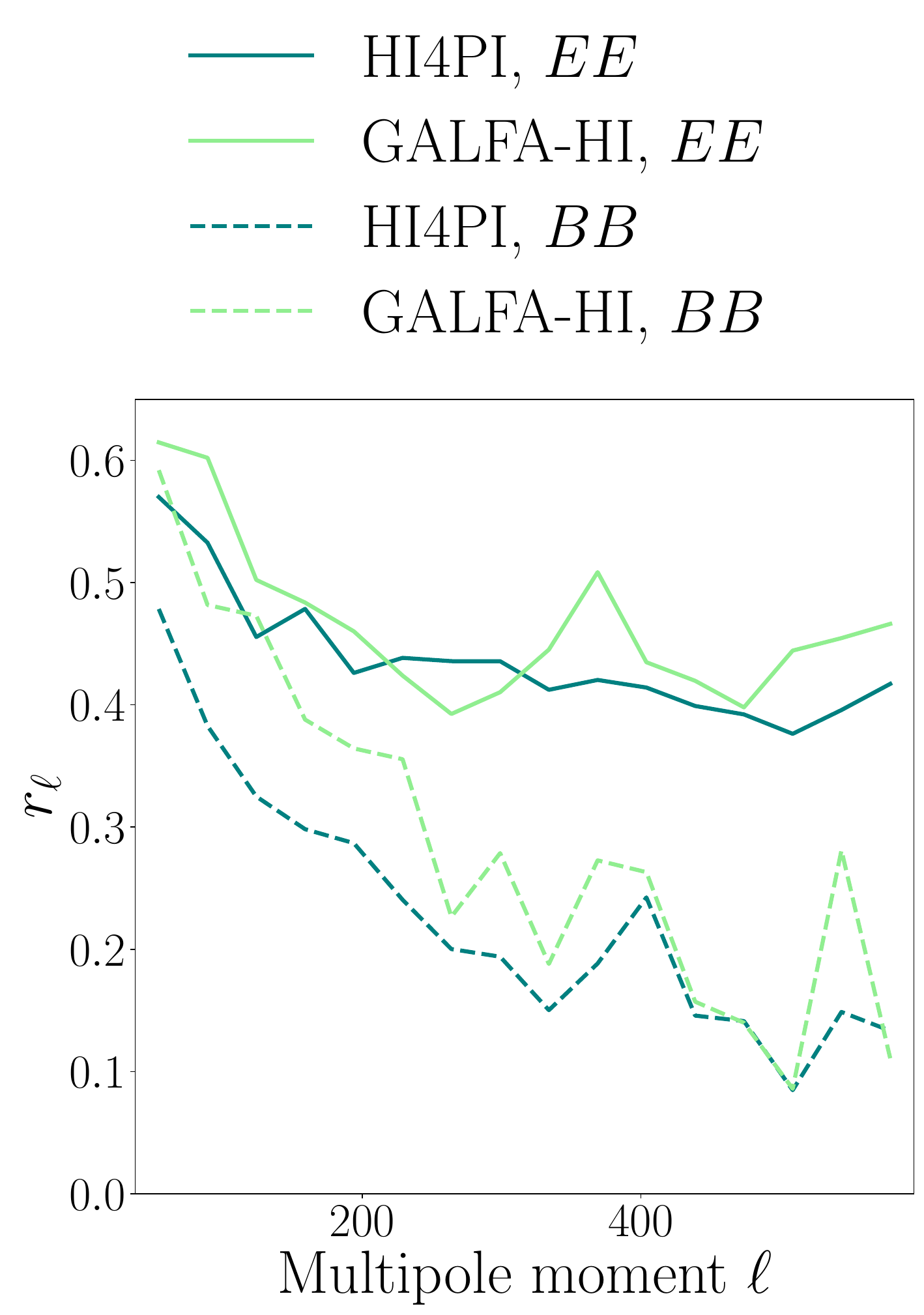}
\caption{Comparison of the~$EE$ (solid) and~$BB$ (dashed) correlation ratios with the Planck \com\ maps at 353~GHz of the~\hip s using GALFA-\hi\ data smoothed to a FWHM of~7$'$ (light green) and~\hipi\ data (teal). The~\hip s are integrated over a similar velocity range and constructed using the Hessian algorithm. The correlations are calculated on a combination of the GALFA-\hi\ mask with the Planck~70\% sky fraction mask. The teal lines are the same as those in \figref{fig:cvp}.
\label{fig:gvh}}
\end{figure}

We compare the~\hip s based on GALFA-\hi\ data smoothed to~FWHM~=~7$'$ and on~\hipi\ data by plotting their~$EE$ and~$BB$ correlation ratios with the Planck \com\ maps at 353~GHz in \figref{fig:gvh}. We use a~$\sim 23$\% sky fraction mask by combining the GALFA-\hi\ mask with the Planck~70\% sky fraction mask. The~\hip\ constructed from the GALFA-\hi\ dataset is more strongly correlated in~$B$ modes with the polarized dust emission than that constructed using the~\hipi\ dataset. The improvement is at the~$\sim~10$\%~level at multipoles~$\ell~<~350$. This implies that higher-resolution~\hi\ data are useful for better modeling the polarized dust foreground in~$B$ modes. If we smooth the GALFA-\hi\ dataset to a resolution of FWHM~=~16$'$ (the resolution of~\hipi) instead of~$7'$, the~$BB$ improvement becomes negligible. 

The~$EE$ and~$BB$ correlation ratios between the~\hip\ constructed from~\hipi\ data and the Planck \com\ maps at 353~GHz do not increase after smoothing the~\hipi\ dataset. For the rest of the analysis in this paper, we use the~\hipi\ dataset at its native resolution.

\section{Spherical Rolling Hough Transform} \label{sec:sphericalrht}
As discussed in the previous section, the Hessian algorithm is sensitive to the local curvature in images and thus is most sensitive to structure at the image resolution. To explore different filament morphologies, however, we need an algorithm with free parameters that help set the scale and shape of the identified filaments. One such algorithm is the RHT \citep{2014ApJ...789...82C,2020ascl.soft03005C} described in \secref{subsec:rht}.

\begin{figure*}[t!]
\includegraphics[width=2\columnwidth]{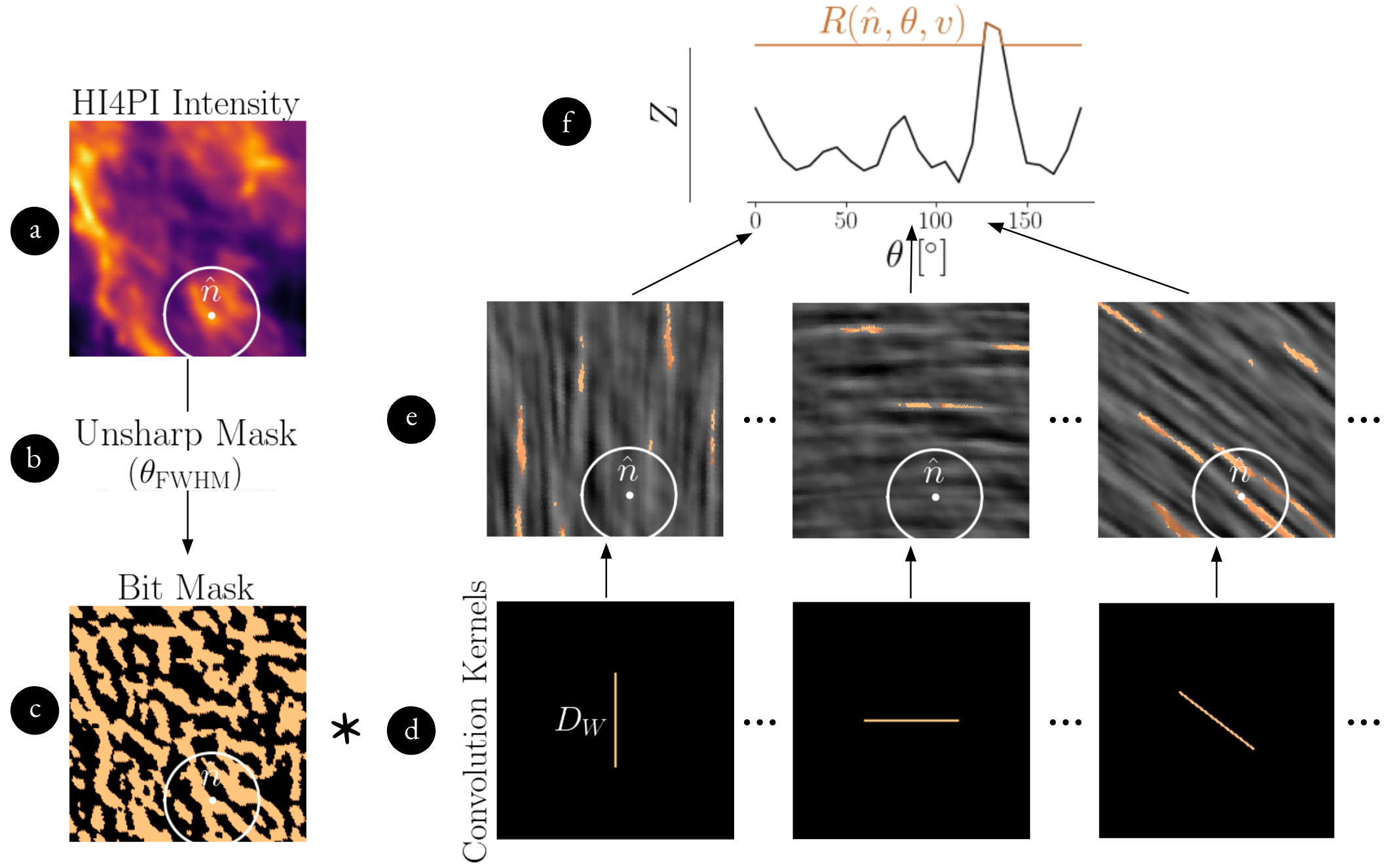}
\caption{Diagram of the Spherical RHT procedure. (a) A flat-sky projection of a~$400'~\times~400'$ patch of sky, centered at~$(l,~b)~=~(15^\circ,~50^\circ)$ and~$v_{\rm lsr}~=~2.03$~\kms\, for the initial~\hi\ intensity channel with width~1.3~\kms\ from the~\hipi\ Survey. The diameter of the white circle drawn around pixel~$\hat{n}$ is equal to~$D_{\rm W}$, the length of the convolution kernels, three of which are shown to scale (d). We chose~$D_W~=~160'$ in this case. (c) The resulting binary map of the preprocessing steps (b; Steps 1 and 2 in \secref{subsec:rht}) with~$\theta_{\rm FWHM}~=~10'$ applied to (a). The convolution kernels (d) are both rotated and convolved with (c) in spherical harmonic space. (e) The results of the convolutions between (c) and (d). (f) The result of the convolutions for pixel~$\hat{n}$ over orientations~$\theta$. A threshold, $Z=0.7$ in this case, is applied to the result of the convolutions (Step 4 in \secref{subsec:rht}), leaving~$R(\hat{n}, \theta, v)$ (copper). The colors in (e) and (f) are set to match, i.e., the pixels (e) have a copper-like color scale where the resulting intensities (f) pass~$Z$ and a gray color scale otherwise.
\label{fig:flowchart}}
\end{figure*}

The code for the RHT algorithm is publicly available and has been applied to a variety of astronomical images, including molecular clouds \citep{2016MNRAS.460.1934M,2016MNRAS.462.1517P}, magnetohydrodynamic simulations \citep{2016ApJ...833...10I}, depolarization canals \citep{2018A&A...615L...3J}, the solar corona \citep{2020ApJ...895..123B}, and supernova remnants \citep{2020ApJ...903....2R}. The algorithm has been adapted to work on resolved stars for stellar stream detection \citep{2022ApJ...926..166P} and extended to add the ability to identify filaments with user-specified widths \citep{Carriere:2022b}.

The RHT algorithm currently runs on flat-sky projections of small patches of the sky. Therefore, to achieve results over the full sky, small patches of the spherical map need to be projected into separate flat-sky images. This is time-consuming and the reprojection step may produce distortion effects. Hence, we implement an alternative algorithm that enables RHT computation directly on the sphere by utilizing spherical harmonic convolutions.

\begin{figure*}[t!]
\includegraphics[width=2\columnwidth]{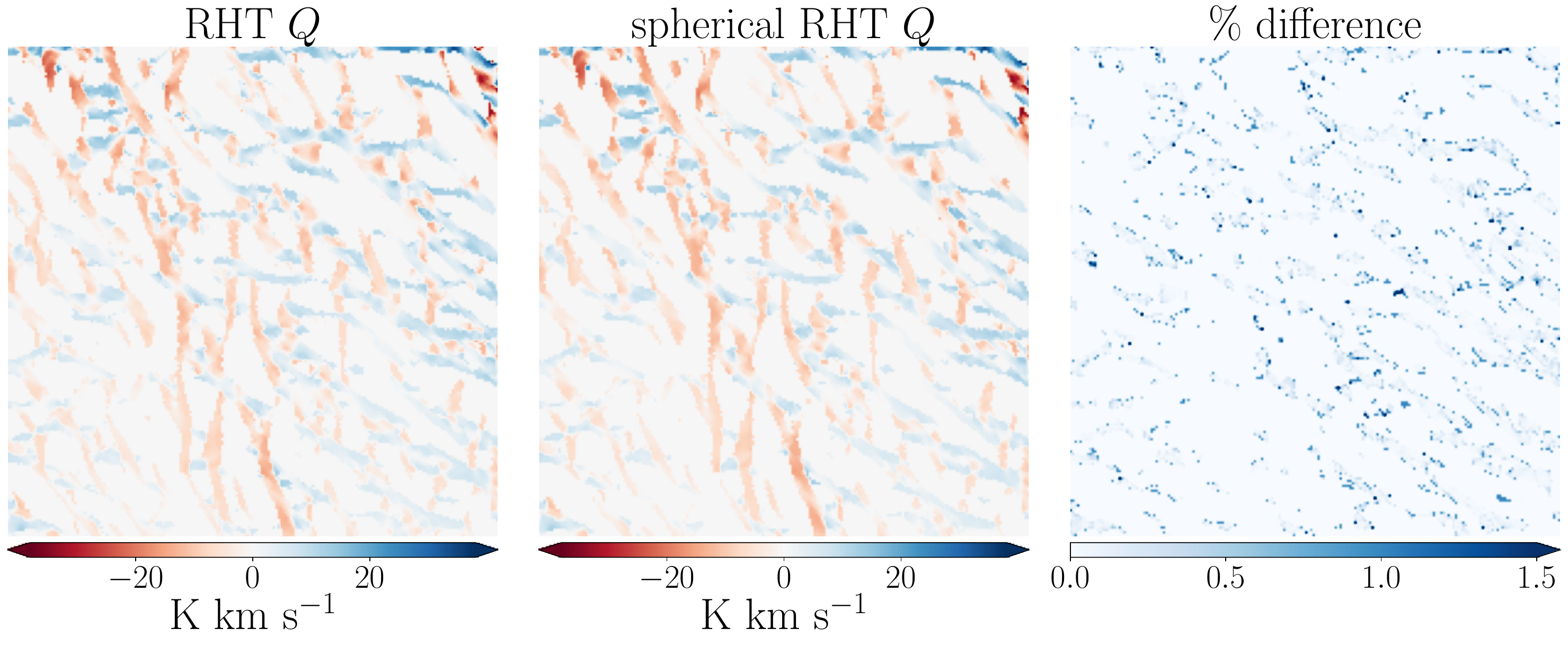}
\caption{Stokes~$Q$ map projections of a~$15^\circ~\times~15^\circ$ patch of sky, centered at $(l,~b)~=~(15^\circ,~50^\circ)$, of~\hip s for one velocity slice of the~\hipi\ Survey centered at~$v_{\rm lsr}~=~2.03$~\kms\ with width 1.3 \kms\ constructed using the RHT (left) and the Spherical RHT (middle) algorithms with parameters~$D_W=75'$,~$\theta_{\rm FWHM}=30'$, and~$Z=0.7$. The map on the left is used in \citet{ClarkHensley}. The map on the right is the percentage difference between the map on the left and the map in the middle.
\label{fig:RHTvSRHT_maps}}
\end{figure*}

\subsection{Spherical Convolutions\label{subsec:conv}}
In a flat geometry, a convolution between two maps can be computed from the product of their Fourier representations. On the sphere, the convolution can be expressed similarly as a product of their spherical harmonic representations. A major difference, however, is that the spherical harmonic representation is weighted by Wigner matrices \citep{Wandelt_2001,2010ApJS..190..267P}. The convolution of a map with spherical harmonics~$a_{\ell m}$ and a convolution kernel defined on the sphere with spherical harmonics~$b_{\ell m'}$ for Euler angles~($\alpha$, $\beta$, $\gamma$) can be written as
\eq{
c(\alpha, \beta, \gamma) = \sum_{m'=-m'_{\rm max}}^{m'_{\rm max}}
\sum_{m=-\ell_{\rm max}}^{\ell_{\rm max}} e^{i m' \alpha} e^{i m \gamma} C_{m'm} (\beta),
}
where
\eq{
C_{m'm} (\beta) \equiv \sum_{\ell = 0}^{\ell_{\rm max}} b^*_{\ell m'} D^{\ell}_{m' m}(\beta) a_{\ell m},
}
and~$D^{\ell}_{m' m}$ are the so-called Wigner matrices. In our case,~$\alpha$ would represent the orientation of the convolution kernel, and~$\beta$ and~$\gamma$ represent the latitude and longitude of the sky, respectively. Refer to \citet{2010ApJS..190..267P} for more details.

The computation of the Wigner matrices is usually the bottleneck of convolution algorithms. This is a major problem in CMB analyses in the context of beam convolutions \citep[e.g.,][]{2000PhRvD..62l3002C}. If the convolution kernel is restricted to a small set of~$m'$ values, then the algorithm can run faster. An example is a symmetric beam, which is restricted to~$m = 0$. Our filamentary kernels are not symmetric, but we can limit the maximum~$m'$ value and retain the intended shape as described in \secref{subsec:alg}. We use \texttt{ducc}\footnote{\url{https://gitlab.mpcdf.mpg.de/mtr/ducc}}, a computationally efficient code for performing convolutions with axially asymmetric convolution kernels.

\subsection{The Algorithm\label{subsec:alg}}
We implement the steps described in \secref{subsec:rht} directly on the sphere rather than on flat-sky image projections. We call this new implementation the spherical Rolling Hough Transform (Spherical RHT). This implementation replaces Step~3 of the RHT algorithm in \secref{subsec:rht} with the spherical harmonic convolutions described in \secref{subsec:conv}. We show a diagram of the full procedure in \figref{fig:flowchart}. This diagram shows how the parameters~$\theta_{\rm FWHM}$,~$D_{\rm W}$, and~$Z$ are used to transform the~\hi\ intensity at each velocity channel~$v$ and pixel~$\hat{n}$ into~$R(\hat{n},\theta,v)$, which is used in Equations~\ref{eq:norm}, \ref{eq:QRHT}, and~\ref{eq:URHT} to construct the~\hi-based Stokes~$Q$ and~$U$ maps for that velocity channel.

\begin{figure*}[t!]
\includegraphics[width=2\columnwidth]{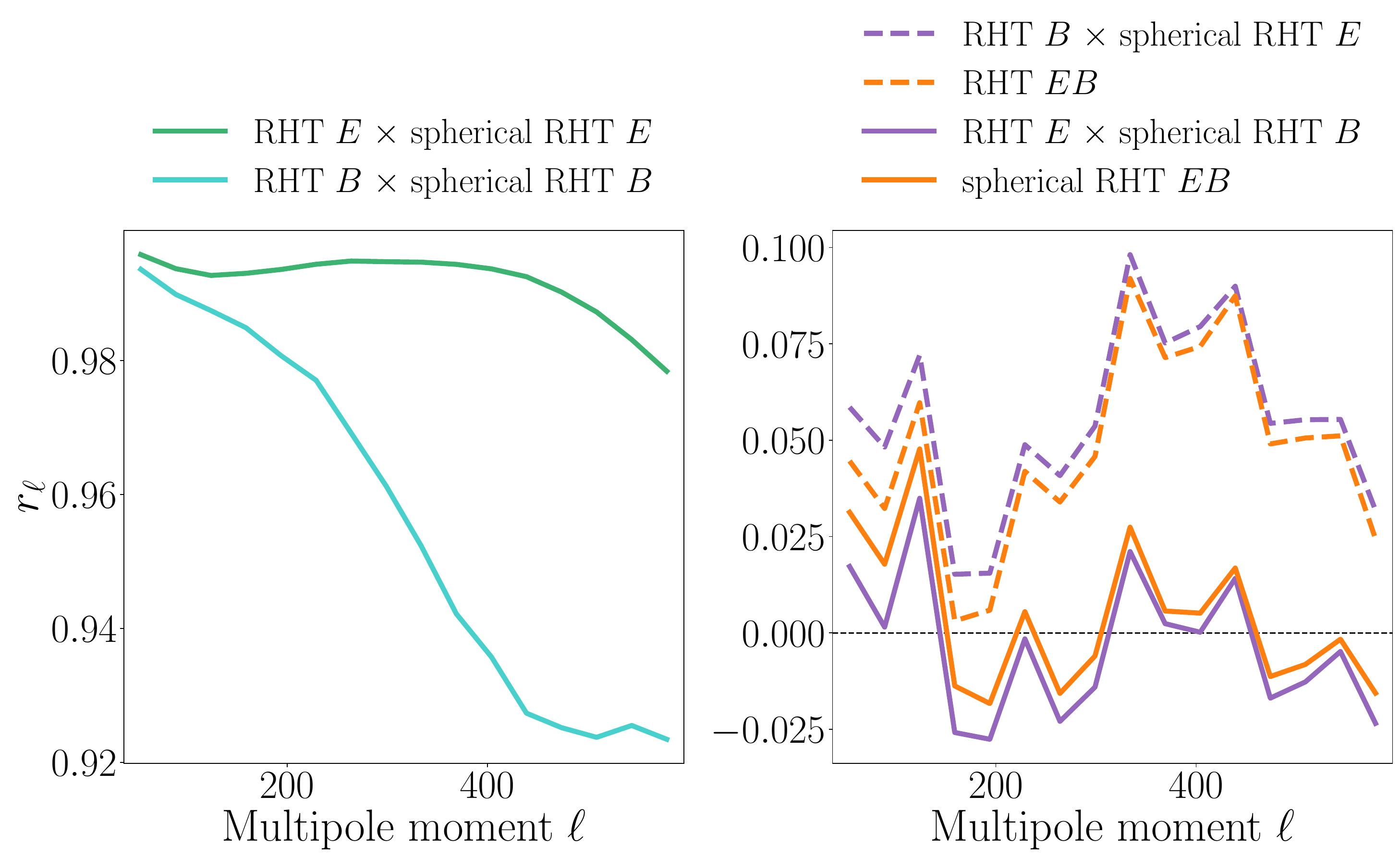}
\caption{Left: the~$EE$ (green) and~$BB$ (blue) correlation ratios between the~\hip\ used in \citet{ClarkHensley} and that reproduced using the Spherical RHT algorithm with the same parameters. Right: the~$EB$ correlation ratios, showing that the likely spurious positive correlation when the~$B$ modes of the \citet{ClarkHensley}~\hip\ are used (dashed) vanish when the curved sky is taken into account and the~$B$ modes produced with the Spherical RHT algorithm are used (solid). The orange (purple) curves represent correlation ratios where the~$E$ and~$B$ modes of the same algorithm (different algorithms) are used. \hipi\ data and the Planck~70\% sky fraction mask are used in these plots. Note the difference in the y-scales between the left and right panels.
\label{fig:rhtxsrht}}
\end{figure*}

We define a convolution kernel as a line of neighboring nonzero pixels of length~$D_{\rm W}$ on a HEALPix grid of a higher resolution than the maps we convolve it with. We then smooth this line of pixels so that the pixelization of the lower-resolution maps captures all of the information in the kernel. This prevents aliasing from small scales that the pixelization of the lower-resolution maps is insensitive to. For instance, when run on \hipi\ data at~$N_{\rm side} = 1024$, we define the kernel at~$N_{\rm side} = 4096$ and smooth it to a FWHM of~3\amin4, the width of a pixel at $N_{\rm side} = 1024$.

We find that limiting the~$m'_{\rm max}$ of the kernel to~50 retains the intended shape of the kernel visually. Also, the results in this paper are identical when~$m'_{\rm max}$ is increased to~100. Limiting the~$m'_{\rm max}$ of the convolution kernel increases the computational efficiency of the algorithm.

We convolve this kernel at different orientations with the map as described in \secref{subsec:conv}. In the standard RHT algorithm, the number of orientations depends on~$D_{\rm W}$. For all~$D_{\rm W}$ used in this paper, dividing the kernel orientations into~25 bins yields consistent results to dividing them into~300 bins. Therefore, we use~25 orientations for computational efficiency. However, the number of orientations is left as a free parameter in the code in case more orientations are necessary for different applications. The code is made publicly available on GitHub\footnote{\url{https://github.com/georgehalal/sphericalrht}} \citep{halal_george_2023_8025777}.

\subsection{Comparison with the RHT\label{subsec:sRHTxRHT}}
We follow the prescription described in \citet{ClarkHensley}, replacing the RHT on small flat-sky projections with the Spherical RHT, to construct full-sky~\hip s and compare the results of the two algorithms.

Once the distribution~$R(\nhat, \theta, v)$ over orientations $\theta$ is obtained for each pixel $\nhat$ at each velocity channel $v$ after Step~4 in \secref{subsec:rht}, \citet{ClarkHensley} construct Stokes~$Q_{\rm HI}$ and~$U_{\rm HI}$ maps as in Equations~(\ref{eq:norm}), (\ref{eq:QRHT}), (\ref{eq:URHT}), (\ref{eq:QRHTsum}), and (\ref{eq:URHTsum}).

\begin{figure}[t!]
\includegraphics[width=1.05\columnwidth]{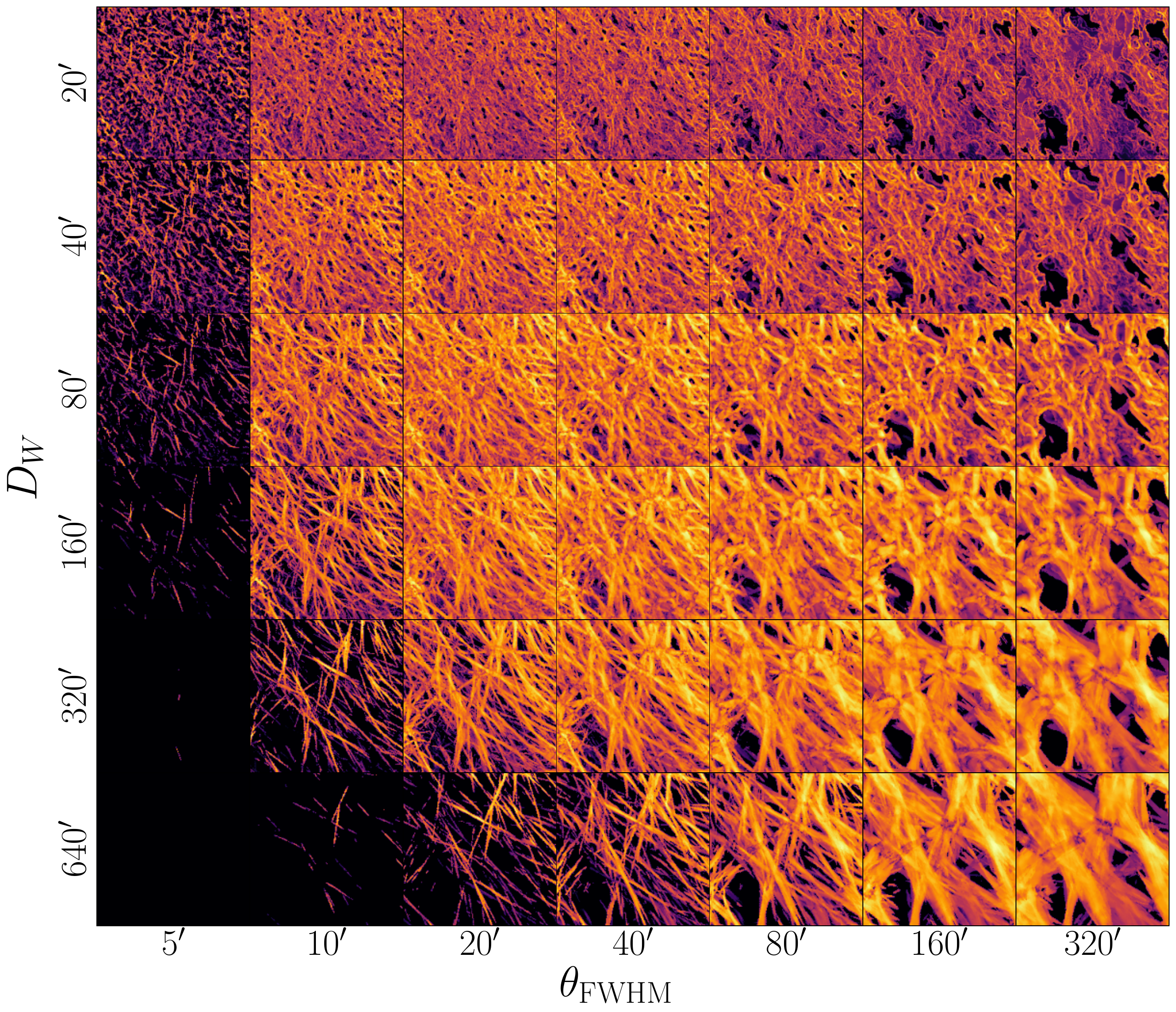}
\caption{Polarized intensity map projections of a~16\adeg7$~\times~$16\adeg7 patch of sky, centered at $(l,~b)~=~(15^\circ,~50^\circ)$, of~\hip s constructed using the Spherical RHT algorithm with different parameters applied to~\hipi\ intensity maps. The parameter~$Z$ is fixed to 0.7 and the Spherical RHT is run on a grid of exponentially increasing parameters between~$5'$ and~$320'$ for~$\theta_{\rm FWHM}$ (to the right) and~$20'$ and~$640'$ for~$D_{\rm W}$ (to the bottom).
\label{fig:paramgridmaps}}
\end{figure}

We run the Spherical RHT on~\hipi\ data with the same velocity binning as in \citet{ClarkHensley} and with the same free parameters,~$\theta_{\rm FWHM}~=~30'$,~$D_W~=~75'$, and~$Z~=~0.7$, and compare the resulting Stokes~$Q$ polarization maps to those of \citet{ClarkHensley} in \figref{fig:RHTvSRHT_maps}. We show flat-sky projections of an example patch of sky of these maps. We also show the percentage difference between the maps and note that the results are qualitatively the same but not numerically identical, and we do not expect them to be. We note that the difference is mostly concentrated at the edges of the filaments. We do not show the Stokes~$U$ results because the conclusions are the same.

To quantitatively test the differences between the two algorithms for constructing~\hip s, we plot the correlation ratio of all different combinations of the~$E$ and~$B$ modes of the two algorithms in \figref{fig:rhtxsrht} using the Planck 70\% sky fraction mask. The correlations between the \citet{ClarkHensley} and the Spherical RHT templates are higher than~98\% in~$E$ modes and higher than~92\% in~$B$ modes across all multipoles considered. Since both algorithms assume no misalignment between the filament orientations and the local magnetic fields as described in \secref{sec:imp}, we expect the~$EB$ correlation to be zero unless the morphology of the filaments across the sky has a preferred chirality as described in \secref{subsec:paramexp}. The~$EB$ correlations are negligible in this figure when the~$B$ modes are predicted by the Spherical RHT algorithm, i.e., when the curved sky is taken into account. However, the~$EB$ correlations are at the~$\sim5\%$ level across the multipole range when the~$B$ modes are predicted by the \citet{ClarkHensley} maps. These results are not latitude dependent. Note that this is not a bug in the RHT code itself. Rather, it is an artifact of projecting each window of the map onto a flat-sky image. While it does not affect any of the results presented in \citet{ClarkHensley}, it is preferable for our morphology investigation that the Spherical RHT does not have this property.

\section{Filament Morphologies} \label{sec:filmorph}
\subsection{Morphological Parameter Space Exploration \label{subsec:paramexp}}
The Spherical RHT enables efficient exploration of the~$D_{\rm W}$,~$\theta_{\rm FWHM}$, and~$Z$ parameter space that governs how the geometry of the filamentary~\hi\ structure is mapped into an~\hip. We use this to investigate what~\hi\ filament morphologies are most predictive of the measured polarized dust emission. This is a continuation of the parameter space exploration performed in~\citet{BKxHI}. That work used the RHT on a small patch that covered~$\sim1\%$ of the sky and assessed the cross-correlation between the~\hip\ and multifrequency polarized dust emission data from BICEP/Keck and Planck. Utilizing the Spherical RHT, we extend this analysis to the full sky in this paper.

\begin{figure*}[t!]
\includegraphics[width=2\columnwidth]{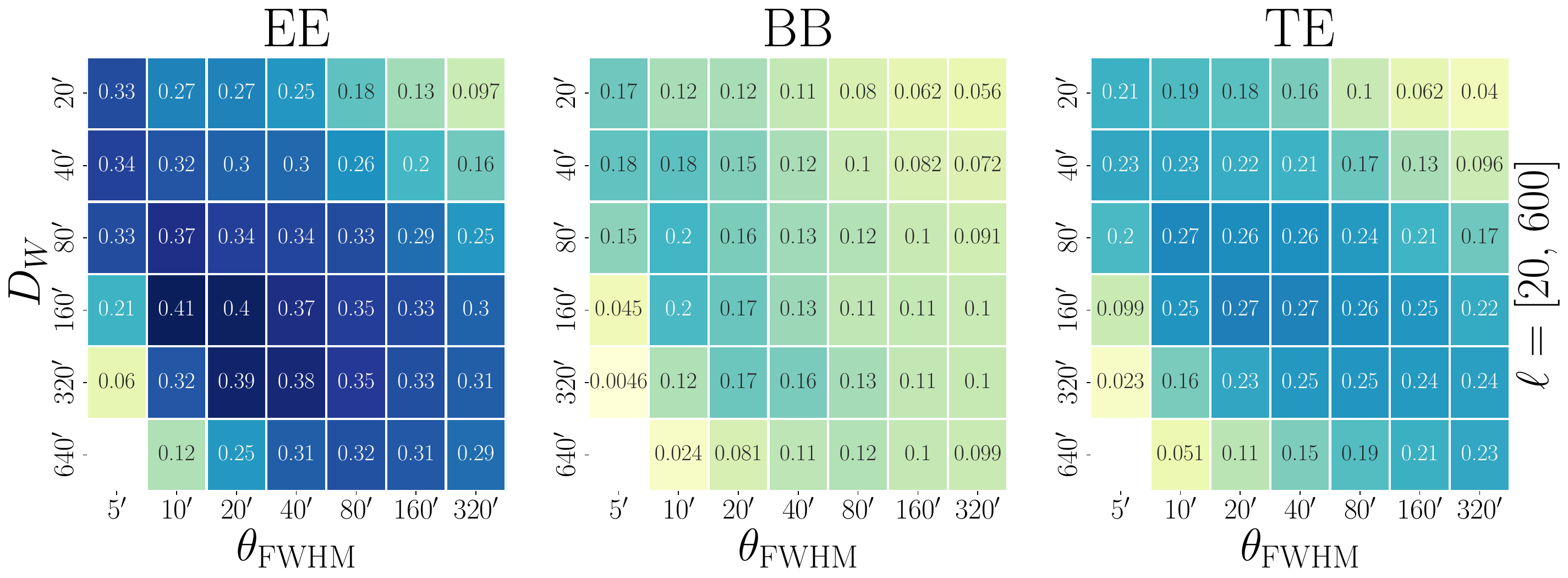}
\caption{The~$EE$ (left), $BB$ (middle), and~$TE$ (right) correlation ratios between the maps in \figref{fig:paramgridmaps} and the Planck \com\ maps over a broadband multipole bin between~$\ell=20$ and~$\ell=600$. The Planck total intensity is correlated with the~$E$ modes of the Spherical RHT-based templates for the~$TE$ case.
\label{fig:paramgridcorr}}
\end{figure*}

Using results from the exploration in~\citet{BKxHI}, we fix the~$Z$ parameter to 0.7, such that the algorithm is only sensitive to structures larger than~70\% of~$D_{\rm W}$. We run the Spherical RHT on a grid of exponentially increasing parameters between~$5'$ and~$320'$ for~$\theta_{\rm FWHM}$ and~$20'$ and~$640'$ for~$D_{\rm W}$. We show the polarized intensity maps corresponding to these parameters in \figref{fig:paramgridmaps}. We calculate the~$EE$, $BB$, and~$TE$ correlation ratios between these maps and the Planck \com\ maps over a broadband multipole bin between~$\ell=20$ and~$\ell=600$ in \figref{fig:paramgridcorr}. For the~$TE$ case, we compute the correlation between the Planck~353~GHz total intensity and the template~$E$ modes.

For the lowest~$\theta_{\rm FWHM}$ and highest~$D_{\rm W}$ case, the RHT intensity is zero because no linear structures cover at least~70\% of~$D_{\rm W}$ after the unsharp mask step removes structure on scales greater than~5$'$. By contrast, for the highest~$\theta_{\rm FWHM}$ and lowest~$D_{\rm W}$ case, the template morphology is more sensitive to lower-intensity diffuse~\hi\ structure, which tends to be less filamentary. This is reflected in the morphology of the~\hip\ in the upper right-hand corner of \figref{fig:paramgridmaps}. At fixed~$\theta_{\rm FWHM}$, the filaments tend to be longer with increasing~$D_{\rm W}$, and at fixed~$D_{\rm W}$, the filaments tend to be wider with increasing~$\theta_{\rm FWHM}$. We find a clear gradient in the correlation coefficient over the parameter space explored with a preference toward~$\theta_{\rm FWHM}\sim10'-20'$, i.e., near the~16$'$~\hipi\ beam scale, and~$D_W\sim80'-160'$. We repeat this exercise for smaller broadband multipole bins and find that the~$\theta_{\rm FWHM}\sim10'-20'$ preferred scale does not change, while the preferred scale for~$D_{\rm W}$ increases slightly when considering larger scales and decreases slightly when considering smaller scales. 

The fact that the preferred~$\theta_{\rm FWHM}$ range is approximately at the beam scale indicates that the thinnest resolved filaments are the most informative about the magnetic field orientation. This means that the~\hi\ filaments that are best-correlated with the polarized dust emission are somewhat thinner and longer than the structures that the \citet{ClarkHensley} analysis was most sensitive to at~$D_W=75'$ and~$\theta_{\rm FWHM} = 30'$. We find that the structures that are more qualitatively filamentary in \figref{fig:paramgridmaps} correlate best in all the metrics in \figref{fig:paramgridcorr}. The less-linear morphologies that populate the upper right-hand portion of \figref{fig:paramgridmaps} correlate poorly with the Planck data. This indicates that elongated linear structures are genuinely the geometry that best describes the polarized dust emission field within the morphological parameter space we can explore.

In the isolated filament case, $E$ modes are primarily sourced along the length of the filament, while $B$ modes are primarily sourced at the edges of the filament \citep{Huffenberger_2020}. Therefore, the~\hip s that maximize the~$EE$ correlation with polarized dust emission are likely the ones that most closely recover the geometry of the dust filaments that dominate the polarized intensity, i.e., their lengths, widths, and orientations. The similarity in the correlation dependence on~$D_{\rm W}$ and~$\theta_{\rm FWHM}$ between the~$EE$ and~$TE$ panels of \figref{fig:paramgridcorr} also supports this conclusion. The~\hip s that maximize the~$BB$ correlation with polarized dust emission are likely the ones whose filaments are at the correct distances from each other, which affects the constructive and destructive interference of the $B$-mode patterns.

\begin{figure*}[t!]
\includegraphics[width=2\columnwidth]{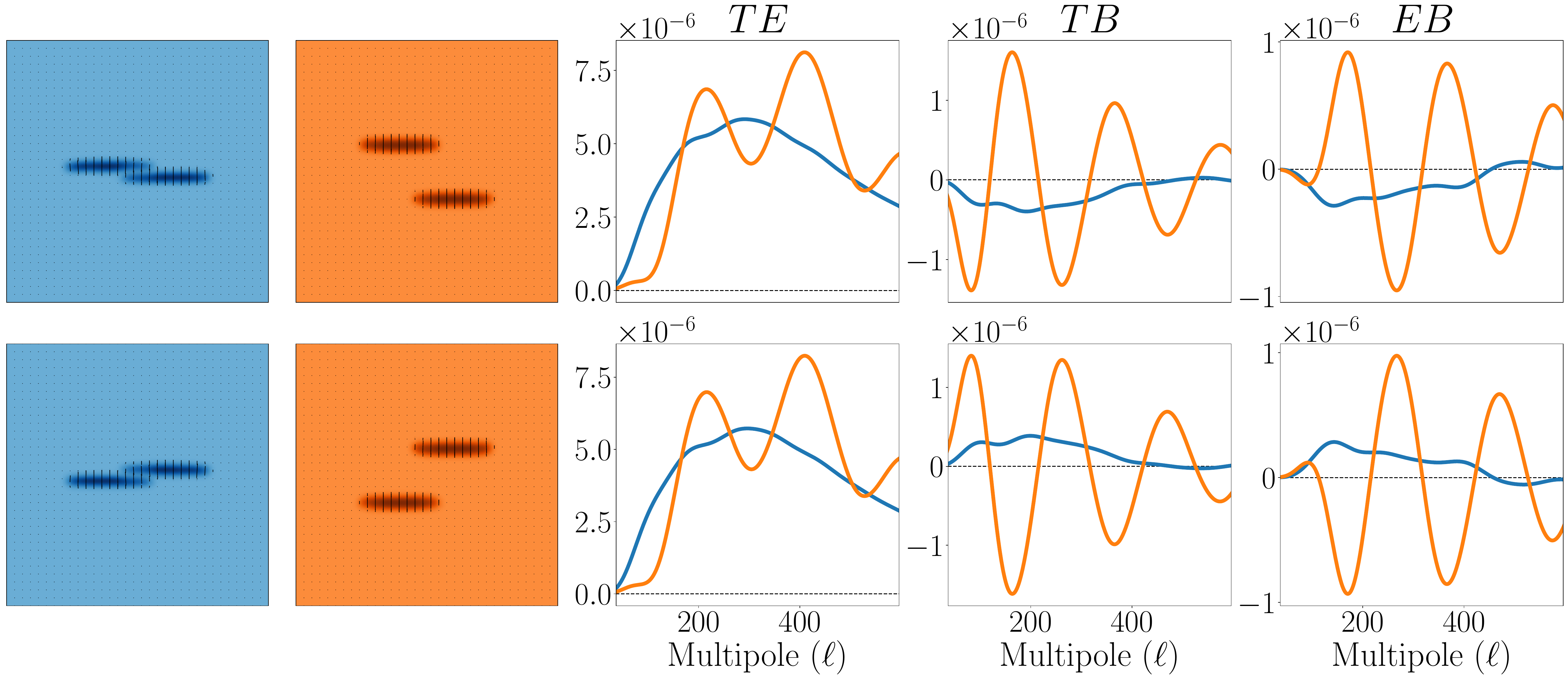}
\caption{Right panel: the~$TE$ (left),~$TB$ (middle), and~$EB$ (right) cross spectra ($D_\ell$) of two synthetic filaments close to each other (blue, left panel) and far from each other (orange, left panel), where the arrangement of the filaments in the top and bottom rows have opposite parities. The headless vectors in the left panels show the polarization angle orientations perpendicular to the lengths of these filaments. The color scale in the left panels represents the polarized intensity.
\label{fig:TBgeom}}
\end{figure*}

To illustrate the~$E$- and~$B$-mode patterns produced in the multifilament case, we run the Spherical RHT algorithm with the parameters used in \citet[][ $D_W~=~75'$,~$\theta_{\rm FWHM}~=~30'$, and~$Z~=~0.7'$]{ClarkHensley}  on maps of two synthetic filaments, slightly offset from one another in longitude. \figref{fig:TBgeom}
shows the results when these filaments are positioned at different distances relative to each other in latitude and when they are reflected relative to each other in longitude. The~$TE$ spectra are positive whether the filaments are close to or far from one another and are unaffected by the parity of the filaments' relative positions. By contrast, the~$TB$ spectra fluctuate around zero when the filaments are far from one another but are only positive or only negative, depending on parity, when the filaments are close to one another. This shows how~$E$ modes mainly depend on the individual filament geometries, whereas~$B$ modes mainly depend on the geometry of the filaments relative to one another. 

\figref{fig:TBgeom} is also a demonstration of how the Spherical RHT algorithm can be used as a tool to extend our intuition about morphological features that produce parity-violating signatures, i.e., nonzero~$TB$ and~$EB$. We show positive-only~$TB$ and~$EB$ signals when the two filaments are positioned close to one another in one handedness and negative-only signals when those filaments are close to one another in the opposite handedness. Characterizing parity-violating signatures in polarized dust emission is interesting for several applications, including confounding cosmic birefringence searches \citep{PhysRevLett.125.221301} and biasing CMB polarization ``self-calibration," which assumes that~$TB$ and~$EB$ signals are due to systematic errors because they must vanish in the standard cosmological model \citep{2016MNRAS.457.1796A}. \citet{Huffenberger_2020} and \citet{Clark2021} have shown that a misalignment between dust filaments and the local magnetic field orientations can produce parity-violating signatures. The~\hip s assume perfect alignment between the filament morphologies and the magnetic field orientations. However, even without local misalignment, chirality in the distribution of filaments or in the non-filamentary polarizing structures could produce these signatures. These may be especially significant on small patches of the sky, whereas the signal may average down if large sky areas show no preferred ``handedness" of the dust intensity distribution.

\subsection{Spherical RHT- and Hessian-based Template Comparison} \label{subsec:hessvsrht}
\begin{figure}[t!]
\includegraphics[width=\columnwidth]{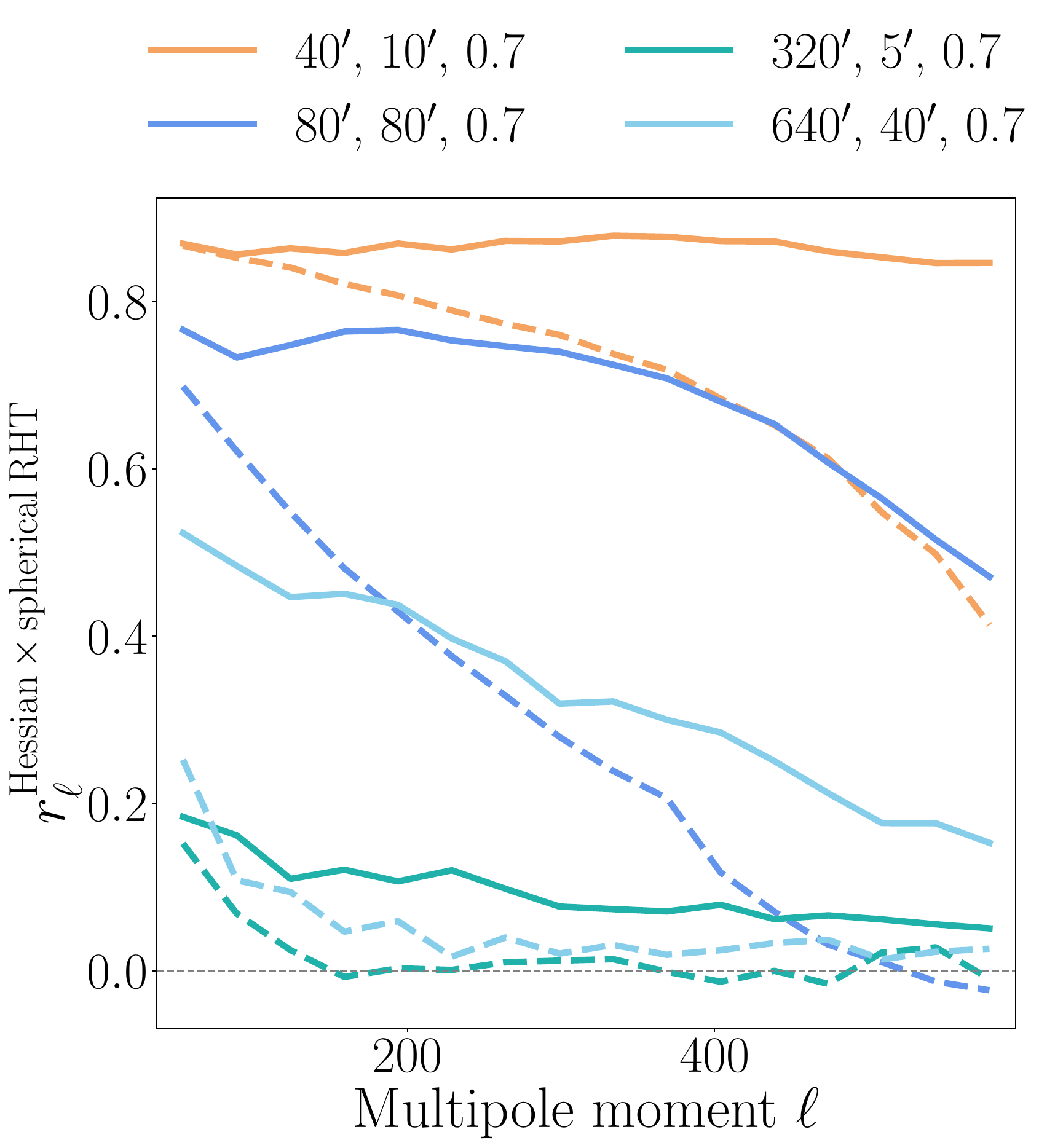}
\caption{The~$EE$ (solid) and~$BB$ (dashed) correlation ratios between the~\hip\ constructed using the Hessian algorithm and four~\hip s constructed using the Spherical RHT algorithm on~\hipi\ data. The Planck~70\% sky fraction  mask was used for the spectra used for calculating these correlation ratios. The parameters listed in the legend are the~$D_{\rm W}$,~$\theta_{\rm FWHM}$, and~$Z$, respectively, defined in \secref{subsec:rht}. The first set of parameters (sandy brown) is the one that correlates the best, and the other three are randomly selected.
\label{fig:EEvBBwbad}}
\end{figure}

We compare the~\hip s produced using the Spherical RHT algorithm to those produced using the Hessian algorithm to understand the factors that affect the correlation of each with the measured polarized dust emission. We start by cross correlating the Hessian-based template with Spherical RHT-based templates constructed using different parameters. We show the~$EE$ and~$BB$ correlation ratios of four of these in \figref{fig:EEvBBwbad}. All of the spectra use the Planck 70\% sky fraction  mask. The~$BB$ correlation ratios are much weaker than the~$EE$ correlation ratios of the same parameters at small scales -- in other words, the Spherical RHT and Hessian templates are more similar to one another in~$E$ modes than in~$B$ modes. While we find a set of Spherical RHT parameters that produces a template that correlates at the~$\sim 90\%$ level in~$E$ modes with the Hessian-based template, none of the parameter sets we test produces a template that correlates higher than~$\sim 60\%$ at~$\ell > 500$ in~$B$ modes. 

\begin{figure*}[t!]
\centering
\includegraphics[width=0.8\textwidth]{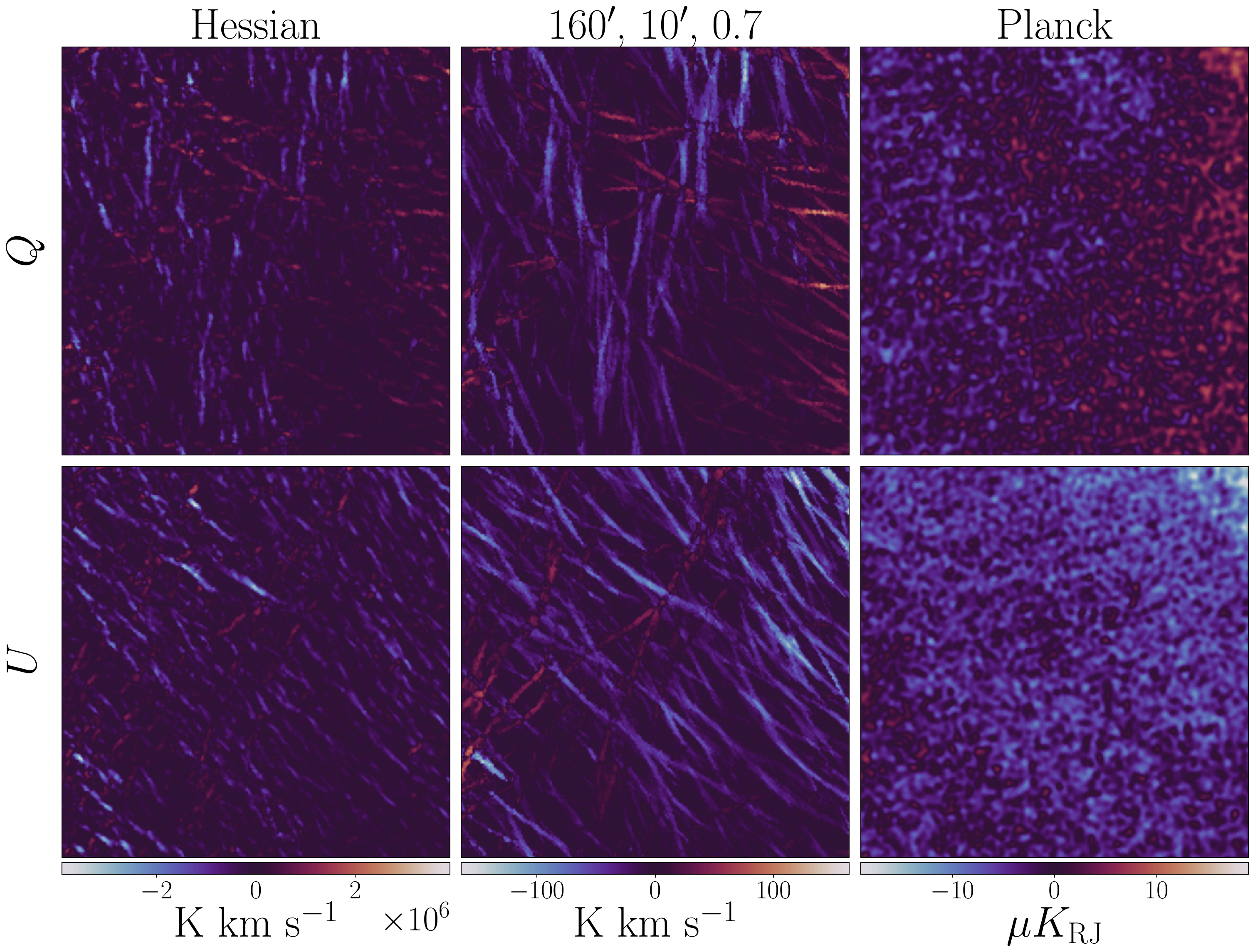}
\caption{Stokes~$Q$ (top) and~$U$ (bottom) map projections of a~16\adeg7~$\times$~16\adeg7 patch of sky, centered at $(l,~b)~=~(15^\circ,~50^\circ)$, of the~\hip s constructed using the Hessian method (left) and Spherical RHT algorithm with parameters~$D_W~=~160'$,~$\theta_{\rm FWHM}~=~10'$, and~$Z~=~0.7$ (middle) applied to~\hipi\ intensity maps, and the full-mission Planck \com\ map (right).
\label{fig:PHSRHT_maps}}
\end{figure*}

We plot the Stokes~$Q$ and~$U$ projections of an example patch of sky for the~\hip s based on the Hessian method and the Spherical RHT algorithm using the best-correlating parameters we found in \secref{subsec:paramexp} and the full-mission Planck~\com\ map in \figref{fig:PHSRHT_maps}. This gives a sense of the different polarization structures predicted by these filamentary templates as compared to the observed total dust field. The Planck maps are noise dominated at small scales. However, the large-scale correspondence is visible in this figure.

\begin{figure*}[t!]
\includegraphics[width=2\columnwidth]{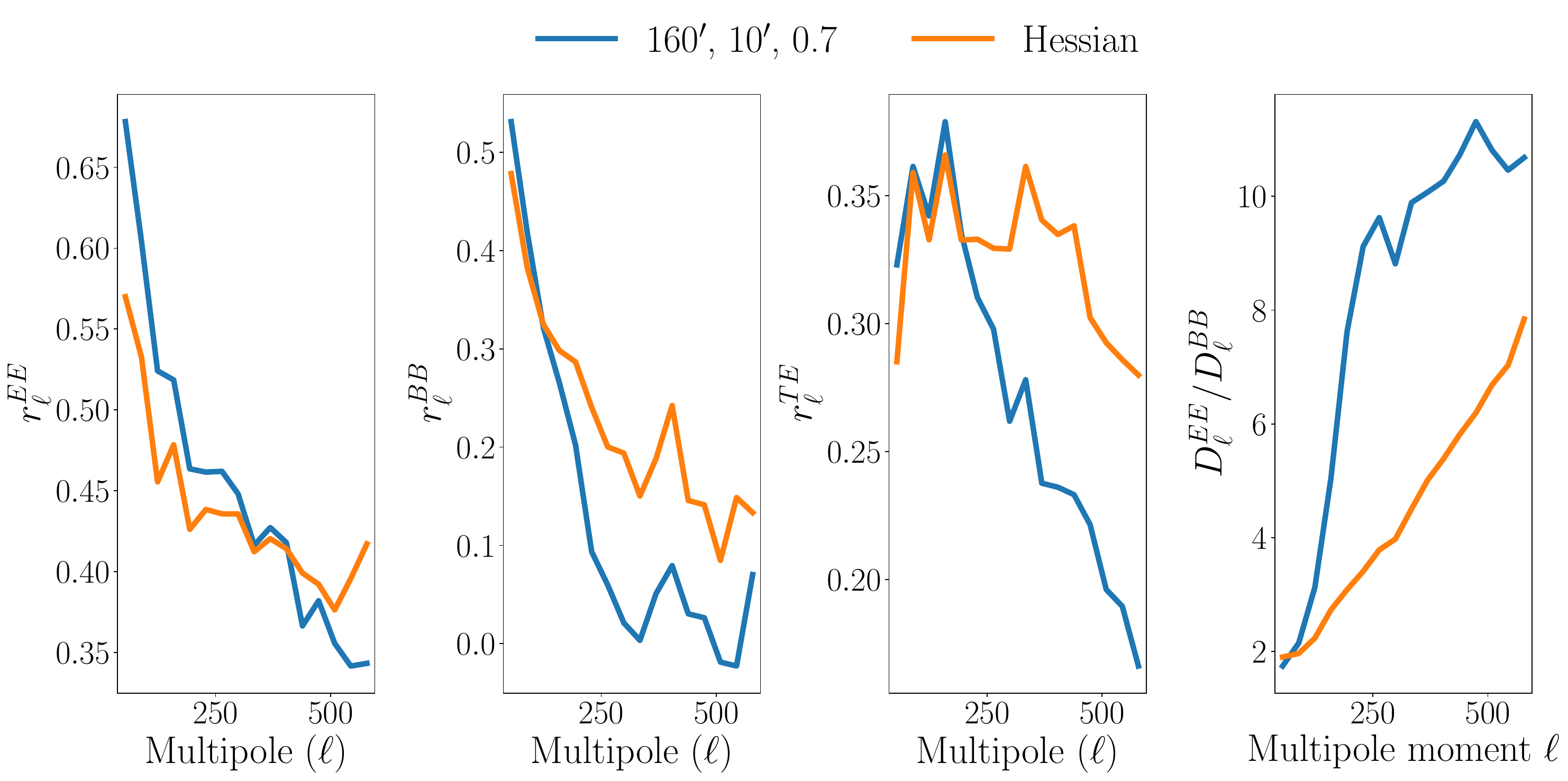}
\caption{The $EE$ (left), $BB$ (middle left), and~$TE$ (middle right) correlation ratios of the Planck \com\ maps with the~\hip\ constructed using the Hessian algorithm (orange) and those constructed using the Spherical RHT with~$D_W~=~160'$,~$\theta_{\rm FWHM}~=~10'$, and~$Z~=~0.7$ (blue). The right panel shows the~$EE$-to-$BB$ ratios of the autospectra of the aforementioned~\hip s. All spectra in this figure are calculated using the Planck 70\% sky fraction mask.
\label{fig:EE_BB_specs}}
\end{figure*}

We compare the correlation of the~\hip s from \figref{fig:PHSRHT_maps} with polarized dust emission. For a broadband multipole bin between~$\ell=20$ and~$\ell=140$ and centered at~$\ell=80$, a range relevant to primordial~$B$-mode detection, we find that the~$EE$ ($BB$) correlation ratios are 0.5 and 0.6 (0.42 and 0.48) for the templates constructed using the Hessian and Spherical RHT algorithms, respectively. We plot the~$EE$,~$BB$, and~$TE$ correlation ratios with the Planck \com\ maps as well as the~$EE$-to-$BB$ autospectra ratio in \figref{fig:EE_BB_specs}. All of the plots in \figref{fig:EE_BB_specs} use the Planck~70\% sky fraction  mask and~\hipi\ data. However, the results are qualitatively similar for different sky fraction masks. 

For the templates constructed with the Spherical RHT algorithm, the~$EE$-to-$BB$ autospectrum ratio peaks at different multipoles for different parameter sets, which determine the typical size of the measured filaments. The peaks are driven by the~$EE$ autospectra in the numerators of these ratios. This is because~$E$ modes predominantly originate along the filaments, so the most sensitive scale of the filament quantification method sets the dominant scale of the~$E$-mode power. Note that these templates correctly predict an excess of~$E$ modes over~$B$ modes, supporting the idea that a preference for filaments in the real sky to be magnetically aligned results in an observed~$EE$-to-$BB$ ratio higher than unity \citep{2015PhRvL.115x1302C,Planck2016XXXVIII}. These templates, however, may overpredict the fraction of power in~$E$ modes on certain scales. Any method of quantifying filament orientations will be more or less sensitive to structure on particular scales \citep[see the discussion in][]{Hacar2022}. This means that an \hip\ built from filament orientations will have \hi-based polarized intensity concentrated at particular angular scales, as is the case in this work. Some of the overprediction of~$E$ modes is likely due to this scale dependence not being representative of the hierarchical filamentary morphology of the real sky. Additionally, the \hip\ may have excess $E$-mode power in part because they only model the filamentary component of the polarized dust emission, whereas the real sky contains additional polarized emission in extended structures that may not resemble filaments. The~$BB$-to-$EE$ ratio of the real polarized dust emission observed by Planck over large sky areas is~$\sim0.53 \pm 0.01$ over the multipole range $40 \leq \ell \leq 600$ \citep{PlanckCollaboration:2020}.

Although we found that the Spherical RHT template correlates better with polarized dust emission than the Hessian template between~$\ell=20$ and~$\ell=140$, the Hessian algorithm correlates better at higher multipoles, especially in~$B$ modes, as shown in \figref{fig:EE_BB_specs}. Because the RHT and Hessian templates predict different distributions of both polarization angles and polarized intensity, we investigate whether one of these factors is driving the stronger correlation between the Hessian-based template and the polarized dust emission data. 

We implement a modification to the~Spherical RHT-based template construction to make its orientation angle selection more similar to that of the Hessian algorithm. While the Hessian algorithm determines the orientation of the filaments based on the local eigenbasis, the~\hip\ constructed with the RHT algorithm computes a mean over all orientations weighted by the result of the convolutions at those orientations $R(\nhat, \theta, v)$; (see \secref{sec:imp}). Therefore, we instead take the angle at the peak of $R(\nhat, \theta, v)$ as the orientation of the filament. We find that for small window diameters~$D_{\rm W}$, this increases the~$E$-mode correlation with the polarized dust emission by~$\sim5\%$. However, the effect is negligible in~$B$ modes and for large window diameters. This behavior is expected because larger window diameters are more likely to have singly peaked~$R(\nhat, \theta, v)$, such that the weighted mean of the angles and the angle at peak~$R(\nhat, \theta, v)$ are similar.

Each Stokes~$Q$/$U$ template is constructed from~\hi-based orientations and~\hi-based polarized intensity. We construct hybrid templates where the orientations are derived from the Spherical RHT and the polarized intensities are derived from the Hessian, and vice versa. Using the Spherical RHT parameters that produce the best correlation with the Hessian algorithm, we construct additional templates using Spherical RHT orientations, but weighting the Stokes~$Q$ and~$U$ maps by the Hessian eigenvalue-based weighting~$w_{\rm H}(\nhat, v)$ described in \secref{subsec:hipcons} instead of the~\hi-intensity-based weighting~$I_{\rm HI}(\nhat, v)$. Similarly, we construct another hybrid template using Hessian-derived orientations and Stokes~$Q$ and~$U$ maps weighted by the~\hi-intensity-based weighting instead of the Hessian eigenvalue-based weighting. The main difference between the two polarized intensity weighting maps is that the Hessian eigenvalue-based one has a more uniform weighting across different filaments than the~\hi-intensity-based one. The standard deviation divided by the mean of the logarithm of the Hessian eigenvalue-based weighting maps is~$\sim0.1$ and~$\sim0.25$ for the~\hi-intensity-based one. Also, for a wide filament with relatively abrupt edges, the Hessian eigenvalue-based weighting upweights those edges relative to the rest of the filament. This is not the case for the~\hi-intensity-based weighting.

\begin{figure*}[t!]
\includegraphics[width=2\columnwidth]{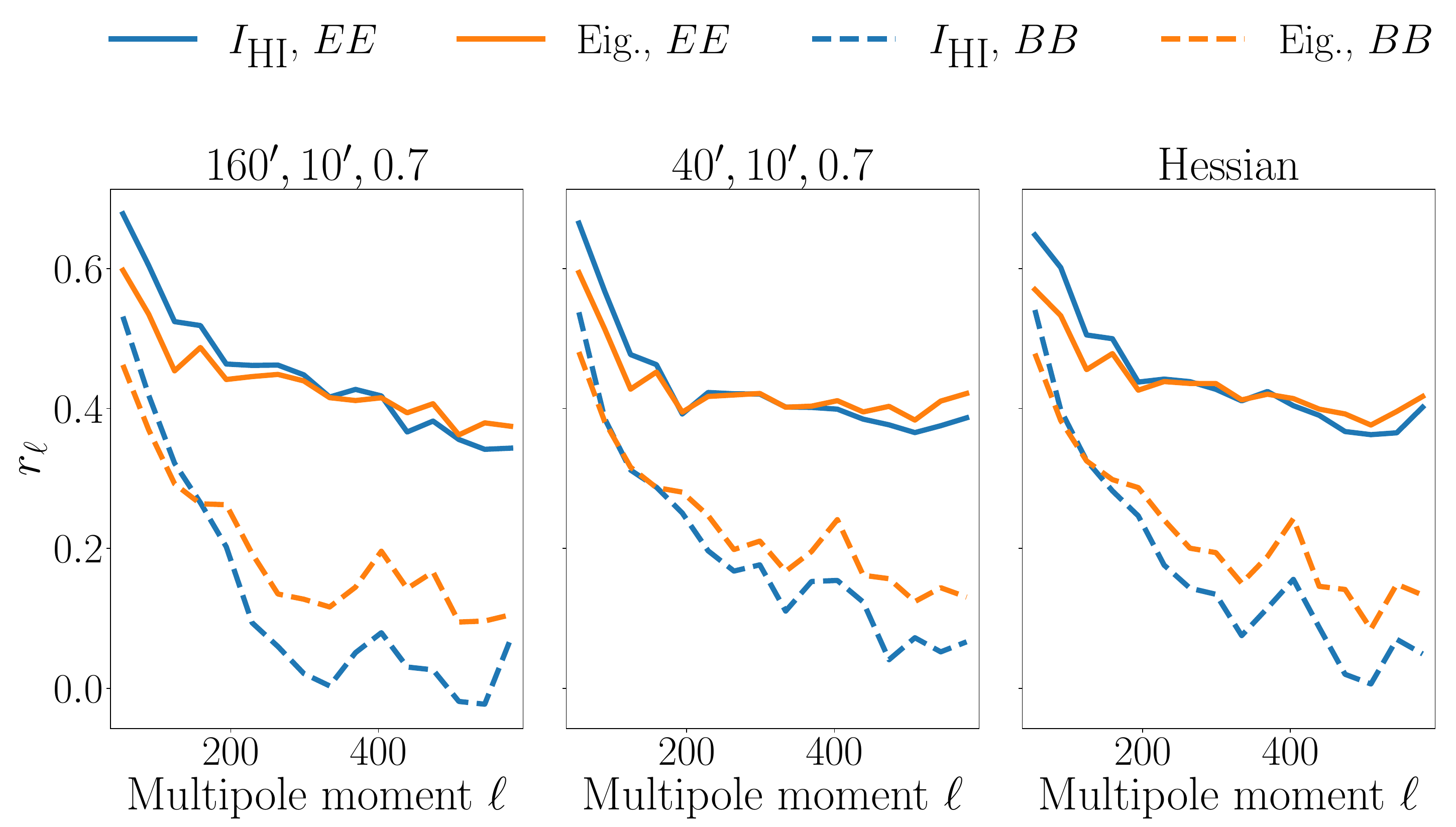}
\caption{The $EE$ (solid) and $BB$ (dashed) correlation ratios of the Planck \com\ maps with the~\hip\ constructed using different algorithms for the orientation angle calculations (different panels) and different weighting schemes (different colors). The left and middle panels use the Spherical RHT algorithm with different parameters for the orientation angle calculations, and the right panel uses the Hessian algorithm for those calculations. The parameters listed in the titles of the left and middle panels are the~$D_{\rm W}$,~$\theta_{\rm FWHM}$, and~$Z$, respectively, defined in \secref{subsec:rht}. In each panel, the~\hi\ intensity-based polarized intensity weighting (blue) is compared with the Hessian eigenvalue-based polarized intensity weighting (orange).
\label{fig:weighting}}
\end{figure*}

We compare the~$EE$ and~$BB$ correlation ratios with the Planck \com\ maps of these hybrid maps compared to those of the maps constructed with the original polarized intensity weighting in \figref{fig:weighting} to isolate the effect of the weighting scheme. We find that the Hessian eigenvalue-based weighting increases the correlation at higher multipoles. This is especially the case for~$B$ modes, where the improvement is at the level of~$\sim10\%$. The improvement is less obvious in~$E$ modes, especially when the Hessian algorithm is used for the orientation angle calculation. In \figref{fig:weighting}, we repeat the exercise with the set of parameters that we found in \figref{fig:EEvBBwbad} to maximize the correlation with the Hessian-based template and confirm the same qualitative conclusion. These results are not latitude dependent. They are also consistent with our results in \figref{fig:EE_BB_specs}, where we see that the improvement in the~$B$-mode correlation at small scales for the Hessian-based template over the Spherical RHT-based template is more obvious than the~$E$-mode correlation. Therefore, we attribute most of the enhancement in the~$B$-mode correlation of the template constructed using the Hessian algorithm to the polarized intensity weighting applied to the different filaments in the Stokes~$Q$ and~$U$ maps. This indicates that the relative weighting of the filaments relative to one another affects~$B$ modes more than it affects~$E$ modes. We correlate the weighting maps directly with the Planck \com\ polarized emission. The correlation with the~\hi-intensity-based weighting map is stronger at large scales and weaker at small scales compared to the correlation with the Hessian eigenvalue-based weighting. This is consistent with the results for the~\hip s themselves.

\section{Conclusions} \label{sec:sum}
We examine the impact of various alterations to~\hip s on the correlation with polarized millimeter-wave observations. This correlation probes the relationship between filamentary ISM structures and the magnetic field, and we investigate what~\hi\ structures are most predictive of the magnetic field orientation. We also use this framework to quantify the contribution of ISM filaments to the polarized dust emission power spectra. This is useful for CMB foreground separation. We make the~\hip s discussed in this work publicly available at \href{https://doi.org/10.7910/DVN/74MEMX}{doi:10.7910/DVN/74MEMX}.

We summarize the conclusions of this work below.
\begin{enumerate}
    \item We improve the~$B$-mode correlation between the Hessian-based template and the polarized dust emission by~$\sim5\%$ over that in \citet{Cukierman} by limiting the~\hi\ velocity range used to~$-13$~\kms~$<~v_{\rm lsr}~<~16$~\kms. This is similar to the LVC range proposed in \citet{2020ApJ...902..120P}. The correlation with dust polarization is worse for wider velocity ranges when using the Hessian method because the Hessian is sensitive to artifacts in low-signal, high-absolute velocity channels.
    
    \item We quantify the correlation between the Hessian-based template and the polarized dust emission in different masks of the sky and find it to be highest (at the~$\sim 30\%-60\%$ level) in the region between the Planck 20\% and 40\% sky fraction masks and lowest (at the~$\sim 10\%-20\%$ level) in the region that covers the 20\% of the sky with the highest integrated dust intensity, i.e., the lowest Galactic latitudes.
    
    \item We introduce the Spherical RHT algorithm, an efficient version of the RHT algorithm that uses spherical harmonic convolutions to run directly on the sphere. We find that the Spherical RHT fixes a spurious~$EB$ signal present at the~$\sim 5\%$ level in the~\citet{ClarkHensley}~\hip\ due to projection effects.

    \item We use the Spherical RHT to explore the parameter space of filament morphologies and their resulting polarization patterns. We find that the thinnest resolved~\hi\ filaments are the most informative for determining the magnetic field orientation. We also find that when using the Hessian method, the~\hip\ constructed from the GALFA-\hi\ data smoothed to~7$'$ correlates~$\sim10\%$~better with the~$B$-mode polarized dust emission field than a template constructed from the~16$'$~\hipi\ data. This motivates the use of even higher resolution \hi\ data, such as the forthcoming Galactic Australian Square Kilometre Array Pathfinder \citep[GASKAP;][]{GASKAP} and the Deep Synoptic Array \citep[DSA-2000;][]{hallinan2019dsa2000}.

    \item We use the Spherical RHT to demonstrate that parity-violating morphologies in the ISM can give rise to nonzero~$TB$ and~$EB$ even when local structures are perfectly aligned with the magnetic field. Since cosmic birefringence could lead to parity-odd polarization signals in the CMB, it is important to quantify parity-odd polarized dust emission as a foreground to those signals.
    
    \item We apply the Spherical RHT to maps of synthetic filaments. We show how individual filament geometries mainly affect the~$E$-mode pattern, whereas the positions and orientations of the filaments relative to one another mainly affect the~$B$-mode pattern. We encourage the reader to use the Spherical RHT for exploring the polarization signatures of other synthetic filament morphologies.
    
    \item We compare the Spherical RHT- and Hessian-based polarization templates. We find that the most significant difference in the correlation with polarized dust emission is in~$B$ modes at small scales, where the Hessian-based template produces a higher correlation. We find that this is due to the difference in the polarized intensity weightings of the templates. The Spherical RHT-based template uses the~\hi\ intensity distribution, while the Hessian-based template uses the Hessian eigenvalue map, which tends to be more uniform. This indicates that~$B$ modes are more sensitive to the polarized intensity weighting of different filaments relative to one another than~$E$ modes are.
\end{enumerate}

The correlation ratio between the integrated~\hipi\ intensity map and the Planck total intensity map at~353~GHz over the Planck~70\% sky fraction Galactic plane mask decreases from~$\sim~80\%$ at $\ell~\sim~40$ to~$\sim~35\%$ at $\ell~\sim~600$ \citep{Cukierman}. This imperfect correlation is partly due to the cosmic infrared background (CIB) in the Planck total intensity map, as well as uncorrelated data noise and systematics. The correlation between the~\hip s and the dust polarization in $EE$ decreases from~$\sim~60\%$ at $\ell~\sim~40$ to~$\sim~45\%$ at $\ell~\sim~600$. The strength of this correlation is partly affected by the fact that the~\hip s do not quantify the diffuse, non-filamentary component of the dust, nor the small local misalignments between the orientations of filaments and magnetic fields.

In addition to providing intuition on the filamentary polarized dust emission patterns, these conclusions provide a step forward in modeling dust polarization using~\hi\ data. This work has focused on comparisons to the observed polarized dust emission using polarization power spectra. Future work could consider additional metrics, including those sensitive to non-Gaussian structures in the dust, e.g., Minkowski functionals \citep{Mantz:2008} or the scattering transform \citep{Mallat:2011}. These techniques have recently been used to quantify structures in dust \citep{2022A&A...668A.122D} and in~\hi\ emission \citep{2023ApJ...947...74L} individually.

\begin{acknowledgments}
This work was supported by the National Science Foundation under grant No. AST-2106607 (PI S.E.C.).

This publication utilizes data from Planck, an ESA science mission funded by ESA Member States, NASA, and Canada.

This work makes use of data from the H\textsc{i}4PI Survey, which is constructed from the Effelsberg-Bonn H~\textsc{i} Survey (EBHIS), made with the 100 m radio telescope of the MPIfR at Effelsberg/Germany, and the Galactic All-Sky Survey (GASS), observed with the Parkes Radio Telescope, part of the Australia Telescope National Facility, which is funded by the Australian Government for operation as a National Facility managed by CSIRO. EBHIS was funded by the Deutsche Forschungsgemein-schaft (DFG) under the grants KE757/7-1 to 7-3.

This work makes use of data from the Galactic Arecibo~$L$-band Feed Array~\hi\ (GALFA-\hi) Survey. It is with the Arecibo 305 m telescope, which is operated by SRI International under a cooperative agreement with the National Science Foundation (AST-1100968), and in alliance with Ana G. Méndez-Universidad Metropolitana and the Universities Space Research Association. The GALFA-\hi\ surveys have been funded by the NSF through grants to Columbia University, the University of Wisconsin, and the University of California.

The computations in this paper were run on the Sherlock cluster, supported by the Stanford Research Computing Center at Stanford University.
\end{acknowledgments}

\vspace{5mm}

\software{astropy \citep{2013A&A...558A..33A,2018AJ....156..123A},
          Healpix\footnote{\url{http://healpix.sourceforge.net/}} \citep{2005ApJ...622..759G},
          healpy \citep{2019JOSS....4.1298Z},
          matplotlib \citep{2007CSE.....9...90H},
          numpy \citep{10.5555/2886196},
          scipy \citep{2020SciPy-NMeth},
          ducc\footnote{\url{https://gitlab.mpcdf.mpg.de/mtr/ducc}},
          pspy\footnote{\url{https://github.com/simonsobs/pspy}} \citep{Louis:2020}
          }

\bibliography{sphericalRHTpaper}{}
\bibliographystyle{aasjournal}

\end{document}